COUNCIL OF EUROPE

CONSEIL DE L'EUROPE

# Discrimination, intelligence artificielle et décisions algorithmiques

**Etude du Prof. Frederik Zuiderveen Borgesius**
**Professeur de Droit, Institut des sciences informatiques et de l'information, Université de Radboud, Nijmegen, et Chercheur, Institut du droit de l'information, Université d'Amsterdam (Pay-Bas)**

# TABLE DES MATIERES

## RESUME

Le présent rapport, préparé à l'intention du Service anti-discrimination du Conseil de l'Europe, porte sur la discrimination issue des décisions algorithmiques et d'autres formes d'intelligence artificielle (IA). Cette dernière permet d'atteindre des buts importants (efficacité, santé, croissance économique, par exemple). Mais elle peut aussi avoir des effets discriminatoires, par exemple lorsque l'apprentissage du système se fonde sur des décisions humaines biaisées.

Des organisations publiques et privées peuvent fonder sur l'IA des décisions lourdes de conséquences pour des personnes. Dans le secteur public, l'IA peut par exemple être utilisée dans la prévention policière (police prédictive) ou dans les décisions de versement de pensions, d'aides au logement ou d'allocations de chômage. Dans le privé, elle peut servir par exemple à sélectionner des candidats à un emploi. Les banques peuvent l'utiliser pour accorder ou non un crédit à un consommateur et fixer le taux d'intérêt correspondant. De plus, un grand nombre de décisions de moindre poids peuvent à elles toutes avoir de larges conséquences. La différenciation des prix par l'IA pourrait par exemple aboutir à ce que certains groupes sociaux paient systématiquement plus cher.

Les principaux instruments juridiques permettant de réduire les risques de discrimination par l'IA sont la réglementation anti-discrimination et la réglementation relative à la protection des données. Effectivement appliquées, elles pourraient contribuer à la lutte contre la discrimination illicite. Les États membres du Conseil de l'Europe, les organes de surveillance du respect des droits de l'homme (comme la Commission européenne contre le racisme et l'intolérance) et les organismes de promotion de l'égalité devraient travailler à ce que le respect des normes anti-discrimination en vigueur soit mieux garanti.

Mais l'IA ouvre également sur de nouvelles formes de différenciations injustifiées (que certains qualifieraient de discriminatoires), que n'envisage pas le droit existant. La plupart des textes anti-discrimination ne s'appliquent qu'à la discrimination fondée sur des caractéristiques protégées, comme la couleur de la peau. Ils ne couvrent pas les nouvelles catégories (non corrélées avec des caractéristiques protégées) que peut créer un système d'IA pour différencier des groupes de population. Ces différenciations n'en pourraient pas moins être injustes, par exemple si elles renforcent des inégalités sociales.

Il sera probablement nécessaire de compléter la réglementation pour protéger la justice et les droits de l'homme dans le domaine de l'IA. Mais il ne convient pas de chercher à réglementer l'ensemble de ce dernier par un dispositif unique, car les applications des systèmes d'IA sont trop variées. Les valeurs sous-jacentes et les problèmes changent d'un secteur à l'autre. Il faudra donc envisager des règles sectorielles. D'autres recherches et débats sont encore nécessaires.

# I. INTRODUCTION

Le présent rapport, rédigé à l'intention du Département anti-discrimination du Conseil de l'Europe, porte sur les risques de discrimination liés aux décisions algorithmiques et à d'autres formes d'intelligence artificielle (IA).

L'IA tend à des buts importants, comme l'efficacité, la santé et la croissance économique. Notre société s'appuie sur elle pour de nombreuses tâches : filtrage des courriels indésirables, planification de la circulation, gestion logistique, reconnaissance de la parole et diagnostic médical, notamment. L'IA et les décisions algorithmiques peuvent paraître rationnelles, neutres et impartiales. Malheureusement, elles peuvent aussi déboucher sur des discriminations injustes et illicites. Comme cela avait été demandé, le rapport étudie les questions ci-dessous.

1. *Dans quels domaines les décisions algorithmiques et d'autres formes d'IA ont-elles des effets discriminatoires ou pourraient en avoir dans un avenir prévisible ?*
2. *Quels garde-fous (et recours) prévoit actuellement le droit en ce qui concerne l'IA, lesquels envisage-t-on ?*
3. *Quelles mesures de réduction des risques de discrimination par l'IA peuvent être recommandées aux organisations qui l'utilisent, aux organismes de promotion de l'égalité des États membres du Conseil de l'Europe, et aux organes chargés de la surveillance du respect des droits de l'homme, comme la Commission européenne contre le racisme et l'intolérance ?*
4. *Par quels dispositifs (juridiques, réglementaires, d'autorégulation) est-il possible de réduire les risques ?*

Le mot « discrimination » est utilisé ici pour désigner la discrimination illicite et réprouvée, par exemple au motif du genre, de la couleur de la peau ou de l'origine raciale[1]**.** C'est le mot « différenciation » qui désigne la discrimination neutre et non réprouvée[2].

Le rapport n'examine qu'un seul risque lié aux décisions algorithmiques et à l'IA : la discrimination. De nombreux aspects de l'IA n'y sont donc pas abordés, comme les systèmes d'armes automatiques, les véhicules autonomes, les bulles de filtrage, la singularité technologique, les monopoles fondés sur les données, le chômage de masse que pourraient susciter l'IA et les robots. Il ne parle pas non plus des questions relatives à la vie privée en rapport avec les énormes volumes de données personnelles recueillies pour alimenter les systèmes d'IA.

Le présent travail est une étude documentaire. En raison de contraintes de longueur, il faut n'y voir qu'un rapide panorama plutôt qu'un examen détaillé de tous les aspects de l'IA, des décisions algorithmiques et de la discrimination. Je tiens à remercier Bodó Balázs, Janneke Gerards, Dick Houtzager, Margot Kaminski, Dariusz Kloza, Gianclaudio Malgieri, Stefan Kulk, Linnet Taylor, Michael Veale, Sandra Wachter et Bendert Zevenbergen de leurs précieuses suggestions.

Le corps du rapport est structuré ainsi : le chapitre II introduit l'intelligence artificielle, les algorithmes décisionnels et quelques autres notions essentielles. Puis il se penche sur les questions évoquées ci-dessus. Le chapitre III repère les domaines dans lesquels l'IA provoque ou peut provoquer des discriminations. Le chapitre IV examine les garde-fous juridiques. Le chapitre V montre comment les organisations peuvent prévenir la discrimination quand elles recourent à l'IA ; il contient aussi des recommandations aux organes de promotion de l'égalité et de surveillance du respect des droits de l'homme sur les façons de réduire les risques de discrimination par l'IA et les décisions algorithmiques. Le chapitre VI propose des améliorations à apporter au cadre juridique, et le chapitre VII présente des conclusions.

---

[1] En conformité avec la tradition juridique, j'utilise ici les termes d'« origine raciale » et de « race », mais je n'accepte pas les théories qui prétendent qu'il existe des races humaines distinctes.

[2] Une décision algorithmique a souvent un but de discrimination (au sens de différenciation ou de distinction) entre des personnes ou des entités. Pour les différents sens du mot « discrimination », voir Lippert-Rasmussen 2014.

## II. INTELLIGENCE ARTIFICIELLE ET DÉCISIONS ALGORITHMIQUES

Les termes d'« IA » et de « décision algorithmique » sont diversement utilisés, et leur définition ne fait pas le consensus. Les paragraphes ci-dessous introduisent rapidement l'intelligence artificielle (IA), les décisions algorithmiques et quelques autres notions connexes.

### *Algorithme*

Un algorithme peut être défini comme la description abstraite et formelle d'un processus de traitement informatique[3]. Ici, « décision » désigne simplement le produit, la conclusion ou le résultat de cette procédure. Pour concrétiser et simplifier les choses, on pourrait assimiler un algorithme à un programme informatique.

La décision algorithmique peut parfois être entièrement automatisée. Un filtre de courriers indésirables, par exemple, peut empêcher de façon complètement automatique les messages indésirables d'arriver dans une boîte. Il peut aussi arriver que des humains prennent des décisions en s'aidant d'un algorithme ; la décision est alors *partiellement* automatisée. Par exemple : un employé de banque peut accorder ou non un crédit à un client sur la base d'une évaluation de sa solvabilité fournie par un système d'IA.

S'agissant de discrimination, les risques des décisions entièrement ou partiellement automatisées sont en grande partie similaires. La recommandation émanant d'un ordinateur peut paraître rationnelle et infaillible, et être suivie aveuglément. Comme l'indiquent Wagner *et al.,* l'être humain peut souvent avoir tendance à valider sans se poser de questions une décision préparée par un algorithme, faute de temps, de compétences ou d'informations contextuelles pour prendre une décision solidement fondée en l'espèce[4]. L'auteur de la décision peut aussi chercher à minimiser sa responsabilité en se conformant à la réponse de l'ordinateur[5]. La tendance à croire l'ordinateur et à suivre son conseil est parfois appelée « biais d'automatisation »[6]. (Nous verrons à la section IV.2 que certaines règles juridiques distinguent les décisions entièrement et partiellement automatisées[7].)

### *Intelligence artificielle*

L'intelligence artificielle (IA) peut être décrite de façon simplifiée comme la science visant à rendre les machines intelligentes[8]. Plus formellement, c'est l'étude et la conception d'agents intelligents[9]. Dans ce contexte, un agent est une chose qui agit, comme un ordinateur[10].

L'IA est un large domaine de recherche, apparu dans les années 1940[11]. Il en existe de nombreux types. Dans les années 1970 et 1980, par exemple, la recherche s'intéressait beaucoup aux « systèmes experts » et aux programmes capables de reproduire l'expertise et les capacités de raisonnement d'un bon spécialiste d'un domaine circonscrit[12]. Les chercheurs ont programmé des ordinateurs capables de répondre à des questions, sur la base de réponses prédéfinies. Ces systèmes experts ont connu un certain succès commercial dans les années 1980[13]. Ils avaient deux inconvénients, observe Alpaydin. D'une part, les règles logiques données au système ne correspondaient pas toujours à la réalité désordonnée du monde. Dans la vie réelle, les choses ne sont pas vraies ou fausses, elles peuvent être plus ou moins vraies : une personne n'est pas simplement vieille ou jeune, la vieillesse augmente progressivement avec l'âge[14]. Et d'autre part, il faut que des experts fournissent le savoir (les réponses) au système : un processus long et coûteux[15].

---

[3] Dourish 2016, p. 3. Voir également Domingos 2015.

[4] Wagner *et al.* 2018, p. 8. Voir également Broeder *et al.* 2017, p. 24-25.

[5] Zarsky 2018, p. 12.

[6] Parasuraman et Manzey 2010. Voir également Citron 2007, p. 1271-1272 ; Rieke, Bogen et Robinson 2018, p. 11.

[7] Se reporter à l'analyse de l'article 22 du GDPR, dans cette même section.

[8] Royal Society 2017, p. 16.

[9] Russel et Norvig 2016, p. 2, citant Poole, Mackworth et Goebel 1998, p. 1 : 'Computational Intelligence is the study of the design of intelligent agents.'

[10] Russel et Norvig 2016, p. p. 4.

[11] Deux des premières publications sont Turing 1951 et McCarthy *et al.* 1955.

[12] Puppe 1993, p. 3.

### *Apprentissage automatique*

Une forme d'IA a connu un vif succès ces dernières décennies : l'apprentissage automatique[16]. Il permet d'éviter que le savoir soit fourni à la machine par des experts. Dans les systèmes de ce type, une tâche est attribuée à la machine, à qui l'on donne aussi un gros volume de données qui lui serviront d'exemples à suivre, ou dans lesquelles elle détectera des régularités. Elle apprend ainsi la meilleure façon de produire le résultat désiré[17].

En première approximation, on pourrait dire que l'apprentissage automatique est un système de prédiction fondé sur des données[18]. Lehr et Ohm proposent une description plus détaillée : c'est un processus automatisé de découverte de corrélations (parfois aussi appelées relations ou modèles) entre les variables d'un jeu de données, souvent dans le but de prédire ou d'estimer un résultat[19].

L'apprentissage automatique s'est beaucoup répandu ces dix dernières années, notamment du fait que l'on dispose de plus en plus de données à fournir aux ordinateurs. Il connaît un tel succès que l'on parle souvent d'IA pour désigner l'apprentissage automatique (qui n'est qu'une forme d'IA)[20].

Exploration de données *(Data Mining)*, mégadonnées *(Big Data)* et profilage sont des notions connexes. L'exploration de données est un type d'apprentissage automatique qui consiste à repérer des constantes intéressantes dans d'énormes quantités de données[21]. C'est aussi une forme de découverte de savoirs dans des données[22]. *Big Data* (ou encore mégadonnées, ou données massives) désigne l'analyse de gros volumes de données[23]. Et le profilage consiste à définir par traitement automatisé de données des profils qui permettront de prendre des décisions concernant des personnes[24].

### *Terminologie utilisée dans le présent rapport*

Le présent rapport sacrifie la précision technique à la lisibilité : les termes d'IA, de système d'IA, de décision d'IA, etc. y sont utilisés sans qu'il soit précisé s'il s'agit d'apprentissage automatique ou d'une autre technique. Un système d'IA peut ainsi désigner, par exemple, un ordinateur qui utilise un algorithme exploitant des données fournies par des opérateurs humains.

---

[13] Alpaydin 2016, p. 51.

[14] Alpaydin 2016, p. 51.

[15] Alpaydin 2016, p. 51.

[16] Alpaydin 2016, p. 51. p. xiii.

[17] Royal Society 2017, p. 19.

[18] Paul, Jolley, et Anthony 2018, p. 6.

[19] Lehr et Ohm 2017, p. 671. Voir également Royal Society 2017, p. 19.

[20] Lipton 2018 ; Jordan 2018.

[21] Han, Pei, et Kamber 2011, p. 33. Voir également Frawley *et al.* 1992, qui définissent l'exploration des données *(data mining)* comme l'extraction significative d'informations implicites, inconnues jusque-là et potentiellement utiles d'un corpus de données.

[22] Han, Pei, et Kamber 2011, p. xxiii. Pour certains, l'exploration de données est une étape dans le processus de découverte de connaissances.

[23] Boyd et Crawford 2012.

[24] Voir Hildebrandt 2008 ; Ferraris *et al.* 2013.

Toujours dans un souci de lisibilité, le rapport parle par exemple d'effets de l'IA un peu comme si cette dernière était une entité capable d'agir par elle-même. Or un système d'IA n'apparaît pas par génération spontanée. Comme l'observent Wagner *et al.,* les constructions mathématiques ou informatiques ne portent pas par elles-mêmes atteinte à des droits de l'homme, c'est leur mise en œuvre et leur application aux actions humaines qui le font[25]. Lorsqu'un système d'IA formule une décision, c'est qu'une organisation a décidé de s'appuyer sur lui pour la prendre.

En fait, une organisation qui commence à recourir à l'IA prend rarement elle-même toutes les décisions sur le système lui-même. Elle pourra déployer un système d'IA dans lequel un grand nombre de choix importants ont déjà été opérés[26]. Dans certains cas, les effets de décisions prises avant même l'achat ou la conception du système d'IA peuvent n'apparaître qu'après la mise en service dans le monde réel. De plus, une organisation se compose de nombreuses personnes : cadres, juristes et informaticiens, par exemple ; mais pour simplifier, le rapport dira parfois qu'une « organisation » fait quelque chose. Le chapitre qui suit examine comment l'IA peut conduire à de la discrimination, et fait ressortir les domaines dans lesquels elle a ou peut avoir des effets discriminatoires.

## III. LES RISQUES DE DISCRIMINATION

***Quels sont les domaines dans lesquels les décisions algorithmiques et d'autres types d'IA ont des effets discriminatoires ou pourraient en avoir dans un avenir prévisible ?***

### 1. COMMENT L'IA PEUT DEBOUCHER SUR DE LA DISCRIMINATION

Cette section examine comment l'IA peut déboucher sur de la discrimination ; la suivante donne des exemples dans lesquels elle l'a fait ou pourrait le faire. Beaucoup de systèmes d'IA sont des « boîtes noires »[27]. Une personne ne comprendra souvent pas pourquoi un système a formulé telle ou telle décision à son sujet. En raison de l'opacité de la décision, il lui sera difficile de voir si elle a été victime de discrimination, par exemple en raison de son origine raciale.

Les décisions fondées sur l'IA peuvent conduire de plusieurs façons à des discriminations. Dans un article fondateur, Barocas et Selbst distinguent cinq façons dont une décision d'IA peut involontairement aboutir à une discrimination[28]. Les problèmes proviennent 1) de la définition de la variable cible et des étiquettes de classe ; 2) de l'étiquetage des données d'apprentissage ; 3) de la collecte des données d'apprentissage ; 4) de la sélection des caractéristiques ; 5) du choix des données indirectes. De plus, 6) les systèmes d'IA peuvent être délibérément utilisés à des fins discriminatoires[29]. Nous allons maintenant revenir sur chacun de ces éléments.

#### *1) Définition de la variable cible et des étiquettes de classe*

L'IA consiste pour l'ordinateur à découvrir des corrélations dans des jeux de données. Une société qui met au point un filtre de courriers indésirables, par exemple, fournit à l'ordinateur des messages étiquetés par des humains comme indésirables ou non. Ces messages étiquetés constituent les données d'apprentissage. L'ordinateur y découvre les caractéristiques d'un courrier indésirable. L'ensemble de corrélations mises au jour est souvent appelé le modèle ou le modèle prédictif. Les messages repérés comme indésirables contiendront souvent, par exemple, certaines expressions (« perte de poids massive », « millions d'euros à gagner », etc.) ou émaneront de certaines adresses IP. Comme le disent Barocas et Selbst, l'algorithme d'apprentissage automatique tire d'exemples pertinents (fraudes précédemment identifiées, courriers indésirables, défauts de paiement, mauvaise santé) les attributs ou les actions (données indirectes) qui peuvent servir à détecter la présence ou l'absence de la qualité ou du résultat recherchés (la variable cible)[30].

---

[25] Wagner *et al.* 2018, p. 8. Voir également Dommering 2006 ; Rieke, Bogen et Robinson 2018, p. 5.

[26] Sur les modes de développement de systèmes numériques modernes, voir Gürses et Van Hoboken 2017. Leur centrage est la protection de la vie privée, mais leur analyse touche également aux systèmes d'IA et à la discrimination.

[27] Pasquale 2015.

[28] Barocas et Selbst 2016. Voir également O'Neil 2016, qui donne une introduction accessible et bien écrite aux risques de discrimination et autres rencontrés dans le domaine des systèmes d'IA.

[29] Barocas et Selbst 2016. Leur regroupement diffère quelque peu.

La variable cible représente ce que l'explorateur de données recherche, expliquent Barocas et Selbst, tandis que les étiquettes de classe répartissent toutes les valeurs possibles de cette variable cible entre des catégories s'excluant les unes les autres[31]. Pour le filtrage des courriers indésirables, par exemple, on s'accorde dans l'ensemble sur les étiquettes de classe  de courrier bienvenu ou indésirable[32]. Mais dans certains cas, la définition de la valeur de la variable cible est moins évidente. Parfois, constatent Barocas et Selbst, il faut créer de *nouvelles* classes pour la décrire[33]. Prenons le cas d'une entreprise qui confie à un système d'IA le soin de classer des réponses à une offre d'emploi pour en extraire de « bons employés ». Comment va-t-on définir le bon employé ? En d'autres termes, quelles devraient être les étiquettes de classe ? Le bon employé est-il celui qui réalise les meilleures ventes ou celui qui n'arrive jamais en retard au travail ?

Certaines variables cibles et étiquettes de classe, expliquent Barocas et Selbst, peuvent avoir un impact négatif plus ou moins marqué sur des classes protégées[34]. Supposons par exemple que les personnes défavorisées habitent rarement en centre-ville et viennent de plus loin que les autres employés pour se rendre à leur travail. Elles seront donc en retard plus souvent, en raison des embouteillages ou de problèmes de transports publics. L'entreprise peut par exemple choisir d'apprécier si un employé est « bon » par l'étiquette de classe « rarement ou souvent en retard ». Mais si les personnes issues de la migration sont en moyenne plus pauvres et habitent plus loin de leur travail, ces étiquettes désavantagent les immigrés, même s'ils font mieux que les autres employés à d'autres égards[35]. En bref, la discrimination peut s'introduire dans un système d'IA en raison de la façon dont une organisation définit les variables cibles et les étiquettes de classe.

### *2) Données d'apprentissage : exemples d'étiquetage*

La décision par IA peut aussi produire des effets discriminatoires si le système « apprend » à partir de données discriminatoires. Barocas et Selbst décrivent deux façons dont des données d'apprentissage biaisées peuvent produire des effets discriminatoires : d'une part, le système peut faire son apprentissage sur des données biaisées ; et d'autre part, des problèmes peuvent surgir si le système apprend à partir d'un échantillon biaisé[36]. Dans les deux cas, le système reproduira le biais.

Les données d'apprentissage peuvent être biaisées si elles reflètent des décisions humaines discriminatoires. C'est ce qui s'est produit dans les années 1980 au Royaume-Uni, dans une école de médecine[37]. L'établissement recevait plus de candidatures qu'il n'avait de places. Il a donc mis au point un logiciel d'aide au tri des candidatures. Les données d'apprentissage étaient constituées par les dossiers d'admission des années précédentes, lorsque des intervenants humains sélectionnaient les candidats. Elles montraient à l'ordinateur les caractéristiques (intrants) corrélées avec le résultat souhaité (admission à l'école de médecine). L'ordinateur a ainsi reproduit ce système de sélection.

---

[30] Barocas et Selbst 2016, p. 678.

[31] Barocas et Selbst 2016, p. 678.

[32] Barocas et Selbst 2016, p. 678-679 (nous avons omis les citations internes).

[33] Barocas et Selbst 2016, p. 679.

[34] Barocas et Selbst 2016, p. 680.

[35] Voir Peck 2013.

[36] Barocas et Selbst 2016, p. 680-681.

[37] Lowry et Macpherson 1988 ; Barocas et Selbst 2016, p. 682.

Il s'est avéré que l'ordinateur défavorisait les femmes et les personnes issues de la migration. Il semblerait que dans les années dont provenaient les données d'apprentissage, les personnes chargées de la sélection des étudiants avaient des préjugés contre les femmes et les personnes issues de la migration. Comme le notait le *British Medical Journal*, le programme n'introduisait pas de nouveau biais, se contentant de reproduire celui qui existait déjà dans le système[38]. Pour conclure, si les données d'apprentissage sont biaisées, il y a des chances pour que le système d'IA reproduise ce biais.

### *3) Données d'apprentissage : collecte des données*

La procédure d'échantillonnage peut aussi être biaisée. Par exemple, dans la collecte de données sur la criminalité, il se pourrait que la police ait interpellé dans le passé plus de personnes issues de la migration. Lum et Isaac observent que si la police se concentre sur certains groupes ethniques et certains quartiers, il est probable que ces catégories seront surreprésentées dans ses fichiers[39].

Si un système d'IA s'appuie sur des données ainsi biaisées, il apprend qu'il est probable que les personnes issues de la migration commettent des infractions. Pour Lum et Isaac, si l'on utilise des données biaisées pour former des modèles prédictifs, ces derniers reproduiront les mêmes biais[40].

Les effets d'un échantillon ainsi biaisé peuvent même être amplifiés par les prédictions de l'IA. Supposons que la police concentre ses activités sur un quartier à forte population immigrée, mais à criminalité moyenne. Elle y enregistra plus d'infractions qu'ailleurs. Comme les chiffres font ressortir un nombre supérieur d'infractions enregistrées (et donc censées s'être produites) dans ce quartier, les autorités y affecteront davantage encore d'agents de police. En d'autres termes, organiser le maintien de l'ordre en se fondant sur des statistiques de criminalité peut créer une boucle de rétroaction positive[41].

Pour prendre un autre exemple, les pauvres peuvent être sous-représentés dans un jeu de données. L'application Street Bump pour smartphone, par exemple, recourt à la géolocalisation pour surveiller l'état des routes dans une ville. Le site explique que des bénévoles l'utilisent sur leur téléphone pour signaler l'état de la route pendant leurs trajets. Ces données sont communiquées en temps réel aux autorités, qui peuvent ainsi procéder aux réparations et planifier leurs investissements à long terme[42]. Si les pauvres sont moins nombreux à posséder un smartphone que les personnes plus aisées, ils seront sous-représentés. Cela pourrait avoir pour effet que les routes détériorées des quartiers pauvres seront moins souvent signalées dans les jeux de données, et donc moins fréquemment réparées. Street Bump était utilisé à Boston, où la municipalité s'efforce de remédier à ce biais dans la collecte des données[43]. Mais cet exemple n'en illustre pas moins que la collecte des données peut produire un biais non voulu dans les données. En bref, un biais dans les données d'apprentissage peut produire un biais dans le système d'IA.

### *4) Sélection des caractéristiques*

Quatrième problème : les caractéristiques (catégories de données) que choisit une organisation pour son système d'IA. Si elle veut utiliser ce dernier pour automatiser une prédiction, il va lui falloir simplifier le monde, pour pouvoir le décrire par des données[44]. Comme le disent Barocas et Selbst, une organisation doit procéder à des choix, pour sélectionner les caractéristiques qu'elle veut observer et introduire dans ses analyses[45].

---

[38] Lowry et Macpherson 1988.

[39] Lum et Isaac 2016, p. 15.

[40] Lum et Isaac 2016, p. 15.

[41] Lum et Isaac 2016, p. 16. Voir également Ferguson 2017 ; Harcourt 2008 ; Robinson et Koepke 2016.

[42] http://www.streetbump.org, consulté le 10 septembre 2018.

[43] Crawford 2013. Voir également Barocas et Selbst 2016, p. 685 ; Federal Trade Commission 2016, p. 27.

[44] Barocas et Selbst 2016, p. 688.

Supposons qu'une organisation veuille sélectionner par prédiction automatisée les candidats qui seront de bons employés. Il est impossible, ou du moins trop coûteux, pour un système d'IA d'évaluer la totalité de chaque dossier de candidature. L'organisation peut alors retenir, par exemple, uniquement certaines caractéristiques applicables à chaque dossier.

Le choix de certaines caractéristiques peut introduire un biais contre certains groupes. Par exemple, de nombreux employeurs aux États-Unis préfèrent les personnes qui ont fait leurs études dans l'une des grandes universités onéreuses. Mais il peut être rare que les membres de certains groupes raciaux fréquentent ces établissements. Le système aura alors des effets discriminatoires dès lors qu'un employeur se fonde sur la fréquentation d'une grande université pour sélectionner les candidats[46]. En bref, une organisation peut susciter des effets discriminatoires par la sélection des caractéristiques qu'utilise le système d'IA dans ses prédictions.

### *5) Données indirectes*

Les données indirectes font aussi problème. Certaines données incluses dans le jeu d'apprentissage peuvent présenter des corrélations avec des caractéristiques protégées. Comme le soulignent Barocas et Selbst, des critères authentiquement pertinents de décisions rationnelles et solidement fondées peuvent aussi constituer des indicateurs fiables d'appartenance à une classe[47].

Supposons qu'une banque utilise un système d'IA censé prédire quels demandeurs de prêt auront du mal à rembourser un crédit. L'apprentissage du système a été fondé sur les données des vingt dernières années, et le jeu ne contient pas d'informations concernant des caractéristiques protégées, comme la couleur de la peau. Le système d'IA apprend que les personnes qui ont un certain code postal ont tendance à ne pas rembourser, et il utilise cette corrélation pour prédire le non-remboursement du crédit. Un critère à première vue neutre (le code postal) sert donc à prédire le défaut de paiement. Mais supposons maintenant qu'il y ait une corrélation entre ce code postal et l'origine raciale. Si la banque prend ses décisions sur la base de cette prédiction et refuse d'accorder des crédits aux habitants de ce quartier, cela fait du tort aux membres d'un certain groupe sur le critère de l'origine raciale.

Barocas et Selbst expliquent que le problème provient de ce que les chercheurs appellent un encodage redondant, c'est-à-dire l'appartenance à une classe protégée encodée dans d'autres données. C'est ce qui se passe lorsqu'une donnée ou certaines valeurs de cette donnée sont étroitement corrélées avec l'appartenance à une classe spécifique protégée[48].

Par exemple : un jeu de données qui ne contient pas de données explicites sur l'orientation sexuelle peut tout de même la dévoiler. Une étude de 2009 a montré que les liens d'« amis » sur Facebook révèlent l'orientation sexuelle par une méthode de prédiction précise de l'orientation sexuelle des utilisateurs de Facebook fondée sur l'analyse de leurs liens. Le pourcentage d'« amis » s'identifiant comme homosexuels serait fortement corrélé avec l'orientation sexuelle de l'utilisateur concerné[49].

Ce problème des données indirectes est délicat. Barocas et Selbst indiquent que les informaticiens ne voient pas très bien comment aborder l'encodage redondant d'un jeu de données. Se contenter de retirer les variables concernées de l'exploration des données supprime fréquemment des critères d'une pertinence démontrable et justifiée dans la décision à prendre[50]. La seule façon de garantir que les décisions ne désavantagent pas systématiquement les membres de catégories protégées est de réduire la précision générale de toutes les déterminations[51].

---

[45] Barocas et Selbst 2016, p. 688.

[46] Barocas et Selbst 2016, p. 689.

[47] Barocas et Selbst 2016, p. 691.

[48] Barocas et Selbst 2016, p. 692. Voir également Dwork *et al.* 2012.

[49] Jernigan et Mistree 2009.

### *6) Discrimination délibérée*

Il faut aussi parler de la discrimination délibérée[52]. Une organisation peut par exemple utiliser volontairement des données indirectes pour pratiquer la discrimination sur le critère de l'origine raciale. Comme l'observent Kroll *et al.,* des préjugés pourraient conduire un responsable à biaiser volontairement les données d'apprentissage ou à choisir certaines données codant indirectement des classes protégées pour obtenir des résultats discriminatoires[53]. Si l'organisation utilise des données indirectes, la discrimination sera plus difficile à détecter qu'une discrimination franche.

Prenons un exemple hypothétique : une organisation pourrait vouloir écarter les femmes enceintes, et cette discrimination serait difficile à détecter. Le distributeur américain Target aurait constitué un score de prédiction de grossesse fondé sur quelque 25 produits en analysant les habitudes d'achat des clientes. Si l'une d'entre elles achetait ces produits, le magasin pouvait savoir avec une bonne certitude qu'elle était enceinte. Target voulait toucher par la publicité des personnes à un moment de leur vie où elles ont tendance à modifier leurs habitudes d'achat. Il voulait donc savoir quand ses clientes allaient avoir un enfant. Ses statisticiens savaient que s'ils arrivaient à les identifier au deuxième mois, ils avaient de bonnes chances de les fidéliser pour des années[54]. Target utilisait la prédiction pour cibler son marketing, mais une organisation pourrait aussi le faire à des fins discriminatoires[55].

En résumé, toute décision d'IA peut produire de la discrimination d'au moins six façons liées 1) à la définition des variables cibles et des étiquettes de classe, 2) à l'étiquetage des données d'apprentissage, 3) à leur collecte, 4) à la sélection des caractéristiques et 5) aux données indirectes — à quoi il faut ajouter 6) l'utilisation délibérée des systèmes d'AI à des fins discriminatoires. L'IA peut aussi conduire à d'autres types de différenciations injustifiées, ou à des erreurs. Nous reviendrons sur ce point au chapitre VI.

## 2. DOMAINES DANS LESQUELS L'IA SUSCITE DES RISQUES DE DISCRIMINATION

La section qui vient donne des exemples de décisions d'IA ayant suscité ou susceptibles de produire des discriminations.

### *Police, prévention de la criminalité*

Commençons par le secteur public. Un exemple célèbre de système d'IA aux effets discriminatoires est le *Correctional Offender Management Profiling for Alternative Sanctions* (COMPAS)[56]. Il est utilisé dans une partie des États-Unis pour prédire les risques de récidive des inculpés. Il est censé aider le juge à apprécier si la personne commettra une nouvelle infraction ou s'il peut ordonner le sursis avec mise à l'épreuve (liberté sous surveillance). Il ne tient pas compte de l'origine raciale ni de la couleur de la peau. Mais les travaux d'Angwin *et al.* (journalistes d'investigation à ProPublica), ont

---

[50] Barocas et Selbst 2016, p. 720.

[51] Barocas et Selbst 2016, p. 721-722.

[52] Barocas et Selbst 2016, p. 692. Voir également Bryson 2017 ; Friedman et Nissenbaum 1996 ; Hacker 2018, p. 1149 ; Kim 2017, p. 884 ; Vetzo, Gerards, et Nehmelman 2018, p. 145.

[53] Kroll *et al.*2016, p. 682.

[54] Duhigg 2012, citant le statisticien de Target. Sur le cas de Target, voir également Siegel 2013, chapitre 2.

[55] Voir Kim 2017, p. 884.

[56] Voir Equivant 2018.

montré en 2016 qu'il contient un biais contre les noirs[57] : ProPublica a donc estimé que :

> COMPAS prédisait bien le risque de récidive dans 61 % des cas, mais que les noirs non récidivistes avaient deux fois plus de chances que les blancs d'avoir été étiquetés comme présentant un risque supérieur, et que le système faisait l'erreur inverse pour les blancs, chez qui les récidivistes avaient beaucoup plus de chances d'avoir été considérés comme présentant un risque inférieur[58].

De plus, observe ProPublica, les inculpés noirs avaient deux fois plus de chances que les blancs d'être classés à tort comme présentant un risque accru de récidive avec violence. Et les récidivistes violents blancs avaient 63 % de chances de plus d'avoir été considérés à tort comme présentant un faible risque de récidive violente que les récidivistes violents noirs[59].

Northpointe, la société dont émane COMPAS, conteste que le système serait biaisé[60]. ProPubllica et Northpointe divergent principalement sur le degré d'impartialité à en attendre[61]. Des statisticiens universitaires ont fait valoir que dans certains cas, des normes différentes d'impartialité sont mathématiquement incompatibles, ce qui a des conséquences sur ce que la prévention de la discrimination pourrait ou devrait être. ProPublica s'inquiétait de ce que l'on pourrait appeler une « disparité de traitement défavorable », faisant que des groupes différents sont différemment affectés par divers types d'erreurs (les membres d'un groupe ayant par exemple plus de chances d'être considéré comme à haut risque alors qu'ils ne commettent ensuite pas d'infraction). Le bon étalonnage est un autre aspect important des scores de risque. Cela veut dire que dans un groupe dont les membres sont considérées comme ayant 80 % de chances de commettre une infraction, 80 % finissent bien par en commettre une. Les chiffres devraient aussi être les mêmes d'un groupe à l'autre (inculpés noir ou blancs, par exemple), faute de quoi le juge devrait interpréter le « haut risque » différemment selon qu'il s'agit d'un inculpé noir ou blanc, ce qui introduit d'autres biais. Les statisticiens pensent que lorsque la propension sous-jacente à la récidive diffère, il est mathématiquement impossible d'obtenir des taux d'erreur identiques[62].

La police recourt parfois aussi à des systèmes d'IA pour planifier ses interventions par prédiction automatisée des auteurs, du moment et des lieux d'infractions[63]. Comme il a été dit ci-dessus, ces systèmes prédictifs peuvent reproduire, voire amplifier, des discriminations existantes.

### *Recrutement d'employés et d'étudiants*

L'IA peut aussi avoir des effets discriminatoires dans le secteur privé. Nous avons vu, par exemple, qu'elle peut servir au recrutement d'employés ou d'étudiants. Comme l'a montré l'exemple de l'école de médecine britannique, un système d'IA peut provoquer des discriminations en raison d'un biais présent dans ses données d'apprentissage. Amazon aurait abandonné un système d'IA pour le recrutement de son personnel en raison d'un biais défavorable aux femmes. Selon Reuters, l'entreprise se serait rendu compte que son nouveau système n'évaluait pas les candidats à des postes d'informaticiens et d'autres emplois techniques de façon non sexiste[64]. Des données historiques avaient enseigné au système d'Amazon que les candidats étaient préférables aux candidates[65].

---

[57] Angwin *et al.* 2016.

[58] Angwin *et al.* 2016.

[59] Larson *et al.* 2016.

[60] Ce paragraphe a été en grande partie rédigé par Michael Veale.

[61] Les discussions entre ProPublica, Nothpointe et des universitaires sur COMPAS sont parfois assez techniques. On en trouvera une bonne synthèse dans Feller *et al.* 2016. Voir également *A shared statement of civil rights concerns* 2018. Pour le point de vue de Northpointe, voir Dieterich, Mendoza et Brennan 2016.

[62] Voir Chouldechova 2017.

[63] Hildebrandt 2014 ; Ferguson 2017 ; Perry et al. 2013 ; Van Brakel et De Hert 2011.

### *Publicité*

L'IA sert au ciblage de la publicité en ligne, très rentable pour certaines entreprises (Facebook et Google, qui figurent parmi les sociétés les mieux cotées, tirent le gros de leurs bénéfices de la publicité en ligne[66]). Ce type de publicité peut avoir des effets discriminatoires. Sweeney a montré en 2013 qu'une recherche sur des noms à consonance afro-américaine attirait sur Google des publicités conduisant à penser que quelqu'un avait un casier judiciaire. Pour les noms à consonance blanche, Google affichait moins de publicités liées à un casier judiciaire. On peut penser que le système d'IA de Google analysait le mode de navigation des utilisateurs et avait hérité d'un biais racial[67].

Datta, Tschantz, et Datta ont simulé des internautes identiques se déclarant hommes ou femmes dans leurs paramètres. Ils ont ensuite analysé les publicités présentées par Google[68]. Pour les internautes se prétendant de sexe masculin, des publicités d'agences de gestion de carrière promettant des salaires élevés apparaissaient plus fréquemment que pour les internautes se présentant comme de sexe féminin — ce qui ferait penser à de la discrimination[69]. L'opacité du système n'avait pas permis aux chercheurs de voir clairement pourquoi les femmes obtenaient moins d'offres d'emplois à forte rémunération : la faible visibilité au sein de l'écosystème publicitaire (Google, annonceurs, sites internet et utilisateurs) les empêchait de déterminer l'origine de ces résultats[70].

Cet exemple montre comment l'opacité d'un système d'IA peut entraver le repérage de la discrimination et de ses causes. Une personne pourrait être victime de discrimination sans le savoir. Dès lors qu'un système d'IA cible les offres d'emploi sur les hommes uniquement, les femmes pourraient ne pas se rendre compte qu'elles ont été exclues de la campagne publicitaire[71].

L'Autorité néerlandaise de protection des données a estimé que Facebook permet de cibler la publicité sur des caractéristiques « sensibles ». Par exemple, des données relatives aux préférences sexuelles ont été utilisées pour afficher des publicités ciblées[72]. Elle indique que Facebook a modifié le système et rendu ce mode de ciblage impossible[73]. À ProPublica, Angwin et Perris ont montré que Facebook laisse les annonceurs exclure des utilisateurs sur le critère de la race. Le système leur permet d'empêcher des publicités de s'afficher pour les noirs, les hispaniques et d'autres « affinités ethniques »[74]. ProPublica a aussi montré que certaines sociétés se servent des possibilités de ciblage de Facebook pour ne diffuser des offres d'emploi qu'auprès de personnes de moins d'un certain âge[75]. Des chercheurs espagnols ont montré pour

---

[64] Dastin 2018.

[65] Dastin 2018.

[66] Fortune 2018. La société mère de Google s'appelle officiellement Alphabet.

[67] Sweeney 2013.

[68] Datta, Tschantz et Datta 2015.

[69] Datta, Tschantz et Datta 2015, p. 93.

[70] Datta, Tschantz et Datta 2015, p. 92 ; Datta *et al.* 2018.

[71] Munoz, Smith et Patil, 2016, p. 9 ; Zuiderveen Borgesius 2015a, chapitre 3, section 3.

[72] Autorité néerlandaise de protection des données 2017 ; Autorité néerlandaise de protection des données 2017a.

[73] Autorité néerlandaise de protection des données 2017.

[74] Angwin et Perris 2016. Voir également Angwin, Tobin et Varner 2017. Dalenberg 2017 analyse l'application de la législation anti-discrimination de l'UE au ciblage publicitaire. Des ONG ont intenté en 2018 un procès à Facebook pour discrimination en se prévalant de la législation des États-Unis sur l'équité dans le logement ; elles accusaient Facebook d'avoir favorisé l'exclusion de femmes, de vétérans handicapés et de mères célibataires de l'audience potentielle d'offres de logements (Bagli 2018).

[75] Angwin, Scheiber et Tobin 2017.

leur part que Facebook caractérise 73 % des utilisateurs de l'UE par des « intérêts sensibles » : Islam, santé reproductive et homosexualité, par exemple [76]. Les annonceurs peuvent cibler leur publicité sur ces critères.

### *Discrimination par les prix*

Un magasin en ligne peut différencier le prix de produits identiques sur la base d'informations qu'il détient sur le consommateur — ce que l'on appelle la différenciation des prix en ligne. Le magasin reconnaît un visiteur sur son site, par exemple par des cookies, et le classe comme sensible ou insensible au prix. Par la différenciation des prix, il cherchera à demander le prix maximum que le client est prêt à payer[77].

Princeton Review, une société américaine de cours particuliers en ligne, modulait ses prix en fonction de la zone : de 6 600 à 8 400 dollars. On peut penser que les frais de fourniture du service étaient les mêmes dans chaque zone, puisque les services sont offerts sur l'internet. Angwin *et al.* ont constaté que cette différenciation des prix pénalisait les personnes d'origine asiatique : les clients habitant dans un quartier à forte densité d'asiatiques avaient 1,8 fois plus de chances de se voir offrir un prix supérieur, quel que soit leur revenu[78]. L'entreprise ne cherchait probablement pas à faire de la discrimination sur le critère de l'origine raciale. Peut-être avait-elle testé différents prix dans différents quartiers, pour s'apercevoir que dans certaines zones, les clients achetaient la même quantité de services, même à un prix supérieur. Quoi qu'il en soit, certains groupes ethniques payaient ainsi davantage.

### *Recherche et analyse d'images*

Les systèmes de recherche d'images peuvent aussi avoir des effets discriminatoires. En 2016, une recherche sur « trois adolescents noirs » a donné dans Google Images des photos d'identité judiciaire, alors que la recherche sur « trois jeunes blancs » a surtout produit des images de jeunes blancs heureux. En réponse à des réactions indignées, Google a indiqué que les résultats reflètent les contenus trouvés sur l'ensemble du web, la fréquence des types d'images et la façon dont elles sont décrites en ligne ; les descriptions déplaisantes de sujets délicats en ligne peuvent parfois ainsi affecter les résultats d'une recherche d'images[79]. On pourrait en effet dire que le système d'IA de Google reflète simplement la société[80]. Mais même si la faute en est à la société plutôt qu'au système, les images ainsi trouvées peuvent teinter l'opinion des gens.

Kay, Matuszek et Munson ont constaté que les résultats de recherches d'images d'activités professionnelles exagèrent légèrement les stéréotypes de genre et représentent le genre minoritaire sous un jour moins professionnel. On constate aussi une légère sous-représentation des femmes[81].

La reconnaissance d'images par les systèmes d'AI suscite d'autres problèmes. Certains logiciels de reconnaissance ont du mal à reconnaître et à analyser les visages non blancs. Un logiciel de détection faciale de Hewlett Packard ne reconnaissait pas les visages à peau foncée comme des visages[82]. Et l'application Google Photos a reconnu sur une photo un couple afro-américain comme des gorilles[83]. Un appareil de photo Nikon demandait constamment aux personnes d'origine asiatique si quelqu'un avait cligné des yeux[84]. Une personne asiatique s'est fait refuser une photo de passeport par un logiciel parce qu'elle aurait fermé les yeux, alors qu'ils étaient ouverts[85]. Buolamwini et Gebru ont constaté que les femmes à la peau foncée constituent le groupe dans lequel les erreurs de classification sont les plus fréquentes (jusqu'à 34,7 %) ; le taux d'erreur maximale pour les hommes à peau claire et de 0,8 %[86]. Peut-être certaines de ces erreurs provenaient-elles simplement du fait que l'apprentissage des systèmes s'était fondé sur des photos d'hommes blancs.

---

[76] Cabañas, Cuevas et Cuevas 2018. Ces centres d'intérêt sont définis comme des « catégories particulières » de données ou des « données sensibles » dans la législation européenne sur la protection des données. Voir article 9 du Règlement général sur la protection des données de l'Union européenne. Sur la législation relative à la protection des données, voir section IV.2.

[77] Zuiderveen Borgesius et Poort 2017.

[78] Angwin, Mattu et Larson 2015 ; Larson, Mattu et Angwin 2015.

[79] Réponse de Google, citée par York 2016.

[80] Allen 2016.

[81] Kay, Matuszek et Munson 2015.

[82] Frucci 2009.

[83] BBC News 2015. Voir également Noble 2018.

### *Outils de traduction*

Le moteur d'intelligence artificielle d'un outil de traduction automatique peut aussi refléter des inégalités et des discriminations. Si l'on donne à traduire en turc l'énoncé « Il est médecin, elle est infirmière » à Google Translate, on obtient « O bir hemşire. O bir doktor » : deux phrases ne comportant pas d'indications de sexe, car le turc ne fait pas la différence entre « il » et « elle ». Si l'on retraduit la traduction en anglais, Google Translate répond : « elle est infirmière, il est médecin ».

Cet exemple est tiré d'une étude de Caliskan, Bryson et Narayanan, qui montre que la machine peut apprendre dans des textes écrits des associations de mots qui reflètent celles qu'apprennent les humains[87]. En d'autres termes, la langue naturelle contient nécessairement des biais humains, et dès lors que la machine apprend sur des corpus linguistiques, elle reproduit inévitablement les mêmes biais[88].

Prates, Avelar et Lamb ont testé douze langues sans genres, comme le hongrois et le chinois, dans Google Translate. Ils ont écrit des phrases comme « il/elle est ingénieur » dans une langue sans genre et demandé la traduction en anglais. Ils ont conclu que Google Translate a fortement tendance à opter pour le masculin par défaut[89]. De plus, cette tendance est non seulement dominante, mais aussi exagérée dans des domaines exposés à des stéréotypes de genre, comme les emplois touchant aux sciences, à la technique, à l'ingénierie et aux mathématiques[90]. En somme, les outils de traduction fondés sur l'IA peuvent produire des résultats qui reflètent des inégalités de genre existantes. Ces résultats pourraient aussi empirer les inégalités par leur influence sur les représentations des lecteurs.

### *Nuancer le risque*

Nous avons vu que les décisions d'IA peuvent avoir des effets discriminatoires, sans que les systèmes d'IA soient nécessairement pires en cela que les humains. Car bien des humains prennent malheureusement aussi des décisions discriminatoires. Dans certains cas, les systèmes d'IA produisent des discriminations qu'ils ont apprises dans des données reflétant des discriminations humaines. Il faut donc bien savoir si l'on évalue une décision d'IA à la lumière de décisions humaines prises dans le monde réel (malheureusement parfois discriminatoires), ou par référence aux décisions idéales qui se prendraient dans un monde exempt de discrimination[91]. Le but doit bien sûr être un monde exempt de discriminations injustes ou illicites.

Cela mis à part, l'IA pourrait aussi servir à mettre au jour les discriminations et inégalités[92]. Prenons le cas d'un système d'IA qui découvre des stéréotypes de genre dans une galerie de photos. Une façon d'interpréter le phénomène est de dire que le système reflète un comportement stéréotypé existant. Il peut ainsi aider à déceler des inégalités existantes, qui resteraient autrement invisibles.

---

[84] Sharp 2009.

[85] Regan 2016.

[86] Buolamwini et Gebru 2018.

[87] Caliskan, Bryson et Narayanan 2017.

[88] Narayanan 2016.

[89] Prates, Avelar et Lamb 2018, p. 1.

[90] Prates, Avelar et Lamb 2018, p. 28.

[91] Voir également Tene et Polonetsky 2017.

[92] Voir Munoz, Smith et Patil 2016, p. 14.

## IV. GARDE-FOUS JURIDIQUES ET RÉGLEMENTAIRES

***Quels garde-fous (et quels recours) prévoit actuellement le droit en ce qui concerne l'IA, et lesquels envisage-t-on ?***

La réglementation anti-discrimination et celle sur la protection des données sont les principales normes susceptibles d'assurer une protection contre la discrimination causée par l'IA. Le présent chapitre aborde successivement ces deux pans du droit ; il examine aussi d'autres domaines envisageables du droit, et l'autorégulation. Il s'agit d'une esquisse, qui s'en tient aux grands principes du droit. Le présent rapport ne saurait traiter par exemple des différences entre les réglementations des États membres du Conseil de l'Europe, la portée territoriale du droit, et la répression des infractions commises par des organisations depuis d'autres États.

### 1. REGLEMENTATION ANTI-DISCRIMINATION

La discrimination est interdite par nombre de traités et de constitutions — notamment la Convention européenne des droits de l'homme[93], dont l'article 14 se lit ainsi :

> « La jouissance des droits et libertés reconnus dans la présente Convention doit être assurée, sans distinction aucune, fondée notamment sur le sexe, la race, la couleur, la langue, la religion, les opinions politiques ou toutes autres opinions, l'origine nationale ou sociale, l'appartenance à une minorité nationale, la fortune, la naissance ou toute autre situation. »[94]

La Convention européenne des droits de l'homme interdit la discrimination *directe comme indirecte*[95]. Nous dirons pour simplifier que la discrimination est directe lorsqu'elle s'appuie sur des caractéristiques protégées, comme l'origine raciale. La Cour européenne des droits de l'homme dit que pour qu'il y ait discrimination directe, « il doit y avoir une différence dans le traitement de personnes placées dans des situations analogues ou comparables », et cette différence de traitement doit être fondée sur une « caractéristique identifiable »[96]. La législation anti-discrimination de l'UE utilise une définition similaire[97].

Toujours en simplifiant quelque peu, nous dirons que la discrimination est indirecte lorsqu'une pratique à première vue neutre se traduit par une discrimination à l'encontre de personnes d'une certaine origine raciale (ou possédant une autre caractéristique protégée)[98]. La discrimination indirecte est désignée par *disparate impact* (disparité

---

[93] Voir par exemple l'article 7 de la Déclaration des droits de l'homme de l'Organisation des Nations Unies, l'article 26 du Pacte international relatif aux droits civils et politiques et l'article 21 de la Charte des droits fondamentaux de l'Union européenne.

[94] Le Protocole nº 12 pose une interdiction similaire, avec un champ d'application plus large à certains égards. « La *jouissance de tout droit prévu par la loi* doit être assurée, sans discrimination aucune, fondée notamment sur le sexe, la race, la couleur, la langue, la religion, les opinions politiques ou toutes autres opinions, l'origine nationale ou sociale, l'appartenance à une minorité nationale, la fortune, la naissance ou toute autre situation. » Article 1, Protocole nº 12 à la Convention de sauvegarde des droits de l'homme et des libertés fondamentales, Série des traités européens – nº 177, Rome, 4.XI.2000. Au 18 septembre 2018, vingt pays étaient parties par ratification ou adhésion au Protocole nº 12. Pour une liste à jour, voir https://www.coe.int/en/web/conventions/search-on-treaties/-/conventions/treaty/177/signatures?p_auth=0Kq9rtcm.

[95] La Convention européenne des droits de l'homme a certains effets horizontaux, mais elle ne couvre pas directement la discrimination dans le secteur privé.

[96] Cour européenne des droits de l'homme, *Biao c. Danemark* (Grande Chambre), nº 38590/10, 24 mai 2016, paragraphe 89.

[97] L'article 2.2.a) de la Directive 2000/43/CE du Conseil du 29 juin 2000 relative à la mise en œuvre du principe de l'égalité de traitement entre les personnes sans distinction de race ou d'origine ethnique dit :« une discrimination directe se produit lorsque, pour des raisons de race ou d'origine ethnique, une personne est traitée de manière moins favorable qu'une autre ne l'est, ne l'a été ou ne le serait dans une situation comparable ». La Directive relative à l'égalité en matière d'emploi et de travail (2000/78/CE), la Directive relative à l'égalité d'accès aux biens et aux services (2004/113/CE) et la refonte de la Directive relative à la mise en œuvre du principe de l'égalité des chances entre hommes et femmes (2006/54/CE) utilisent des définitions semblables. Mais même au sein de l'Union européenne, les législations anti-discrimination ne sont que partiellement harmonisées.

[98] Sur la notion de discrimination indirecte, voir d'une manière générale Tobler 2005 ; Ellis et Watson 2012, p. 148-155.

d'impact) aux États-Unis d'Amérique. Elle est ainsi définie par la Cour européenne des droits de l'homme :

> « […] une différence de traitement [peut] consister en un effet préjudiciable disproportionné occasionné par une politique ou une mesure qui, bien que formulée de manière neutre, opère une discrimination à l'égard d'un groupe […]. Une telle situation s'analyse en une "discrimination indirecte" qui n'exige pas nécessairement qu'il y ait une intention discriminatoire. »[99]

Le droit européen en donne une définition similaire :

> « Une discrimination indirecte se produit lorsqu'une disposition, un critère ou une pratique apparemment neutre est susceptible d'entraîner un désavantage particulier pour des personnes d'une race ou d'une origine ethnique donnée par rapport à d'autres personnes, à moins que cette disposition, ce critère ou cette pratique ne soit objectivement justifié par un objectif légitime et que les moyens de réaliser cet objectif ne soient appropriés et nécessaires. »[100]

Toute décision peut se traduire par une discrimination indirecte involontaire. Le législateur retient donc les effets pratiques de la discrimination indirecte plutôt que l'intention de discrimination de son auteur[101], qui n'est pas pertinente.

La législation anti-discrimination peut servir à lutter contre des décisions discriminatoires d'IA. Par exemple, des décisions de ce type ayant pour effet que les personnes d'une certaine origine ethnique paient plus cher des biens et des services pourraient enfreindre l'interdiction de la discrimination indirecte. Les décisions d'IA peuvent probablement aboutir plus souvent à des discriminations indirectes accidentelles qu'intentionnelles.

La législation anti-discrimination n'en présente pas moins plusieurs points faibles en ce qui concerne les décisions d'IA. L'interdiction de la discrimination indirecte ne constitue pas une règle claire, aisément applicable[102]. La notion elle-même débouche sur des normes assez vagues, souvent difficiles à mettre en œuvre dans la pratique. Il faut pouvoir prouver qu'une règle, une pratique ou une décision apparemment neutre affecte de manière disproportionnée un groupe protégé et constitue donc une présomption de discrimination. Dans bien des cas, on recourt à des preuves statistiques pour démontrer l'effet disproportionné[103].

La Cour européenne des droits de l'homme accepte qu'une accusation de discrimination indirecte puisse être rejetée si son auteur allègue une justification objective :

> « Une politique ou une mesure générale qui ont des effets préjudiciables disproportionnés sur un groupe de personnes peuvent être considérées comme discriminatoires même si elles ne visent pas spécifiquement ce groupe et s'il n'y a pas d'intention discriminatoire. Il n'en va toutefois ainsi que si cette

---

99 Cour européenne des droits de l'homme, *Biao c. Danemark (Grande Chambre),* nº 38590/10, 24 mai 2016, paragraphe 103.

100 Article 2.2.b) de la Directive sur l'égalité raciale 2000/43/CE (typographie et ponctuation adaptées).

101 Cour européenne des droits de l'homme, *Biao c. Danemark* (Grande Chambre), nº 38590/10, 24 mai 2016, paragraphe 103. Voir également Hacker 2018, p. 1153.

102 Nous dirions que l'interdiction de la discrimination indirecte est plutôt un standard qu'une norme. Voir Sunstein 1995 ; Baldwin, Cave et Lodge 2011, chapitre 14.

103 Cour européenne des droits de l'homme, *D.H. et autres c. République tchèque* (Grande Chambre), nº 57325/00, 13 novembre 2007, paragraphes 187-188.

> politique ou cette mesure manquent de justification “objective et raisonnable” »[104].

La justification doit être objective et raisonnable, critères que ne remplit pas une mesure, une pratique ou une règle si :

> « elle ne poursuit pas un but légitime ou si fait défaut un rapport raisonnable de proportionnalité entre les moyens employés et le but visé »[105].

Dans la même veine, le droit de l’UE dit qu’une pratique ne constitue pas une discrimination indirecte pour autant qu’elle soit « objectivement justifié[e] par un but légitime et que les moyens pour parvenir à ce but soient appropriés et nécessaires »[106]. La possibilité pour l’auteur d’une discrimination de se prévaloir d’une telle justification objective dépend des circonstances de l’espèce et appelle un test de proportionnalité nuancé[107]. On ne discerne donc pas toujours clairement si une pratique enfreint l’interdiction de la discrimination indirecte.

La nécessité de démontrer la présomption de discrimination peut aussi susciter des difficultés, ce type de discrimination pouvant ne pas apparaître au grand jour. Prenons le cas d’une personne qui demande un crédit sur le site internet d’une banque. Cette dernière utilise un système d’IA pour accepter ou rejeter les demandes de ce type. Si elle refuse automatiquement le crédit sur son site internet, le client ne voit pas le motif du refus. De plus, il ne saura pas si le système d’IA de la banque refuse les crédits à un nombre disproportionné de femmes, par exemple[108]. Même s’il sait que la décision a été prise par un système d’IA et non pas par un employé, il lui sera difficile de déterminer si elle est discriminatoire.

La notion de caractéristique protégée sur laquelle se fonde la législation anti-discrimination constitue un autre point faible. Les textes sur la non-discrimination mentionnent en général la discrimination (directe et indirecte) fondée sur des caractéristiques protégées, comme la race, le genre ou l’orientation sexuelle[109]. Mais bien des formes nouvelles de différenciations opérées par l’IA paraissent injustes et problématiques, voire discriminatoires, et n’entrent pas dans le champ d’application de la plupart des lois anti-discrimination. Ces dernières présentent donc des lacunes. Nous reviendrons à la section IV.3 sur les différenciations injustes que pourraient ne pas couvrir les textes de lutte contre la discrimination.

Les lois anti-discrimination font obstacle à de nombreux effets discriminatoires de l’IA, en particulier par la notion de discrimination indirecte. Mais elles sont délicates d’application, et présentent des points faibles. La section qui suit examine la réglementation relative à la protection des données.

## 2. REGLEMENTATION RELATIVE A LA PROTECTION DES DONNEES

Le cadre juridique de la protection des données est un outil de défense de la justice et des droits fondamentaux, comme le droit à la vie privée et le droit à la non-discrimination[110]. Il confère des droits à la personne dont les données font l’objet d’un traitement (personne concernée)[111] et impose des obligations aux parties associées au traitement (responsables du traitement)[112]. Il s'appuie sur huit grands principes, que l'on peut résumer comme suit :

(a) des données personnelles ne peuvent être traitées que licitement, loyalement et avec transparence (licéité, loyauté et transparence) ;

(b) ces données ne peuvent être recueillies que dans un but précisé par avance, et ne doivent pas être utilisées à d'autres fins (limitation des finalités) ;

(c) elles devraient être limitées à ce qui est nécessaire en relation avec les finalités pour lesquelles elles sont traitées (minimisation des données) ;

(d) elles devraient être suffisamment exactes et à jour (exactitude) ;

(e) elles ne devraient pas être conservées pendant une durée excessive (limitation de la conservation) ;

(f) elles devraient être traitées d'une manière qui garantisse une sécurité appropriée contre un traitement non autorisé ou illicite, etc. (intégrité et confidentialité)[113] ;

(g) le responsable du traitement est responsable du respect des règles (responsabilité)[114].

---

104 Cour européenne des droits de l’homme, *Biao c. Danemark* (Grande Chambre), nº 38590/10, 24 mai 2016, paragraphe 91 et 92 (nous avons supprimé les citations internes et leurs appels de notes).

105 Cour européenne des droits de l’homme, *Biao c. Danemark* (Grande Chambre), nº 38590/10, 24 mai 2016, paragraphe 90. Voir également Cour européenne des droits de l’homme, *Affaire relative à certains aspects du régime linguistique de l’enseignement en Belgique*, nº 1474/62 et autres, 23 juillet 1968, paragraphe B.10.

106 Article 2.2.b) de la Directive sur l’égalité raciale 2000/43/CE.

107 Collins et Khaitan 2018, p. 21 ; Hacker 2018, p. 1161-1170.

108 Voir Larson *et al.* 2017 pour un exemple réel similaire : *‘These are the job ads you can’t see on Facebook if you’re older’*.

109 Gerards 2007 ; Khaitan 2015.

110 Voir article 1(2) et considérants 71, 75, et 85 du GDPR, et article 1 de la Convention du Conseil de l’Europe sur la protection des données 2018 ; Conseil de l’Europe, Lignes directrices sur la protection des personnes à l’égard du traitement des données à caractère personnel à l’ère des mégadonnées, 2017, paragraphe 2.3.

Ces principes figurent dans la Convention 108 du Conseil de l'Europe sur la protection des personnes à l'égard du traitement automatisé des données à caractère personnel (révisée en 2018)[115] et dans le Règlement général de l'Union européenne sur la protection des données (GDPR, de 2016). On les retrouve dans plus d'une centaine de législations nationales sur la protection des données[116].

Le droit relatif à la protection des données pourrait contribuer à la réduction des risques de discriminations injustes et illicites[117]. Il exige par exemple la transparence du traitement des données personnelles. Les organisations doivent ainsi fournir, par exemple dans une déclaration de confidentialité, des informations sur toutes les étapes de la préparation de la décision d'IA impliquant des données personnelles[118]. Certes, la plupart des gens ne lisent pas les déclarations de confidentialité[119]. Mais ces dernières pourraient être utiles à des chercheurs, à des journalistes et aux organes de contrôle. Si une déclaration donne à penser qu'un traitement pourrait avoir des effets discriminatoires, les autorités peuvent enquêter.

Dans certains cas, le GDPR et la Convention 108 sur la protection des données exigent des organisations (responsables des données) qu'elles procèdent à une analyse d'impact relative à la protection des données, que l'on peut décrire ainsi :

---

111 Article 4(1) GDPR ; article 2(a) Convention du Conseil de l'Europe sur la protection des données 2018.

112 Article 4(7) GDPR ; article 2(d) Convention du Conseil de l'Europe sur la protection des données 2018.

113 Article 5(1)(a)-5(1)(f) GDPR ; articles 5, 7, et 10 Convention du Conseil de l'Europe sur la protection des données 2018.

114 Article 5(2) du GDPR ; article 10(1) Convention du Conseil de l'Europe sur la protection des données 2018.

115 Articles 5, 7, et 10 Convention du Conseil de l'Europe sur la protection des données 2018.

116 Greenleaf 2017.

117 Sur les interactions entre le droit relatif à la protection des données et le droit anti-discrimination, voir Schreurs *et al.* 2008 ; Gellert *et al.* 2013 ; Hacker 2018 ; Lammerant, De Hert, Blok 2017.

118 Article 5(1)(a) ; article 13 ; article 14 GDPR ; articles 5(4)(a) et 8 Convention du Conseil de l'Europe sur la protection des données 2018.

119 Zuiderveen Borgesius 2015.

> l'analyse d'impact est un outil qui sert à l'analyse des conséquences possibles d'une action pour un ou plusieurs points susceptibles d'affecter la société dès lors que ladite action peut présenter des dangers sur ces points ; elle constitue une aide à la prise d'une décision éclairée sur la réalisation de l'action concernée et de ses conditions de déploiement, en fin de compte dans un but de protection[120].

Le GDPR exige une analyse d'impact dès lors que « des opérations de traitement sont susceptibles d'engendrer un risque élevé pour les droits et libertés des personnes physiques », surtout s'il y a recours à des technologies nouvelles[121]. Dans certains cas, il exige systématiquement une analyse d'impact (le risque étant jugé élevé), par exemple lorsqu'une organisation s'appuie sur des décisions entièrement automatisées produisant des effets juridiques ou similaires à l'égard de personnes physiques[122]. Ce qui veut dire qu'il impose une analyse d'impact pour de nombreux systèmes d'IA formulant des décisions qui affectent des personnes physiques[123]. L'analyse doit aussi envisager le risque de discrimination injuste ou illicite[124].

La Convention 108 du Conseil de l'Europe sur la protection des données et la Charte des droits fondamentaux de l'Union européenne prévoient que chaque État membre doit se doter d'une autorité indépendante de protection des données[125]. Cette autorité doit posséder des pouvoirs d'investigation[126]. C'est le GDPR qui donne le plus de détails sur ce dernier point. Une autorité indépendante peut, par exemple, obtenir l'accès aux locaux du responsable du traitement, mener des enquêtes sous forme d'audits sur la protection des données, et ordonner au responsable du traitement de lui communiquer toute information dont elle a besoin et de lui ouvrir l'accès à ses systèmes de traitement des données[127].

### *Règles relatives aux décisions automatisées*

Le GDPR contient des règles spécifiques pour certaines « prises de décision individuelle automatisée »[128]. Ces règles visent notamment à réduire le risque de discrimination illicite[129]. La Convention du Conseil de l'Europe sur la protection des données contient également des règles applicables aux décisions automatisées, moins détaillées toutefois que celle du GDPR[130]. Nous nous concentrerons ici sur ce dernier.

L'article 22 du GDPR (parfois appelé « disposition Kafka ») pose l'interdiction de principe des décisions entièrement automatisées ayant des effets juridiques ou similaires ; il s'applique par exemple au recrutement entièrement automatisé, sans intervention humaine[131]. Le texte qui avait précédé le GDPR contenait une disposition

---

[120] Kloza *et al.* 2017, p. 1. Voir également Groupe de travail (Working Group) « article 29 » 2017 (WP248) ; Binns 2017 ; Mantelero 2017 ; Wright et De Hert 2012.

[121] Article 25(1) GDPR.

[122] Article 35(3)(a) GDPR. Voir également considérant 91 GDPR.

[123] L'article 35(3)(b) et 35 (3)(c) du GDPR pourrait également s'appliquer à certains systèmes d'IA.

[124] Groupe de travail (Working Group) « article 29 » 2017 (WP248), p. 6, p. 14. Voir également Kaminski 2018a, p. 25 ; Edwards et Veale 2017.

[125] Article 8(3) de la Charte des droits fondamentaux de l'Union européenne. Voir également article 51 GDPR ; chapitre IV Convention du Conseil de l'Europe sur la protection des données.

[126] Chapter VI GDPR ; chapitre IV Convention du Conseil de l'Europe sur la protection des données.

[127] Article 58(1) GDPR. L'autorité de protection des données peut également exercer ces droits à l'encontre de sous-traitants (organisations assurant le traitement des données pour le responsable de données).

[128] Article 22 GDPR. L'analyse des règles du GDPR sur les décisions automatisées se fonde sur Zuiderveen Borgesius et Poort 2017, et leur reprend certaines phrases.

[129] Voir considérant 71 GDPR.

[130] Article 9(1)(a) Convention du Conseil de l'Europe sur la protection des données 2018.

[131] Considérant 71 GDPR.

similaire, qui n'avait guère été appliquée dans la pratique[132]. La règle principale de la disposition du GDPR sur les prises de décisions individuelles automatisées se lit ainsi :

> La personne concernée a le droit de ne pas faire l'objet d'une décision fondée exclusivement sur un traitement automatisé, y compris le profilage[133], produisant des effets juridiques la concernant ou l'affectant de manière significative de façon similaire[134].

En résumé, une personne ne doit pas faire l'objet d'une décision entièrement automatisée lourde de conséquences. Le GDPR dit que la personne « a le droit de ne pas faire l'objet » de certaines décisions. Mais on s'accorde à reconnaître en général que ce droit implique l'interdiction de principe de ces décisions[135].

Selon Mendoza et Bygrave, quatre conditions doivent être remplies pour que cette disposition s'applique : a) l'existence d'une décision b) entièrement fondée sur c) le traitement automatisé de données et d) qui a des effets significatifs, juridiques ou similaires, pour la personne[136].

Pourraient par exemple avoir des effets juridiques une décision de justice ou une décision concernant le versement de prestations sociales garanties par la loi, comme une pension de retraite[137]. Pourrait par exemple avoir « des effets similaires » la décision automatisée d'une banque refusant une demande de crédit[138]. Les autorités de protection des données disent également que la différenciation des prix en ligne pourrait aussi affecter quelqu'un « de manière significative » si elle se traduit par des prix excessivement hauts, interdisant à la personne concernée l'accès à certains biens ou services[139].

Il existe des exceptions : l'interdiction de certaines décisions automatisées ne s'applique pas si la décision automatisée : a) est prise avec le consentement explicite de la personne concernée ; b) est nécessitée par un contrat entre la personne concernée et le responsable des données ; c) est autorisée par la loi[140].

Si le responsable des données peut se prévaloir a) du consentement de la personne concernée ou b) de l'exception contractuelle, c'est une autre règle qui s'applique : il doit mettre en œuvre « des mesures appropriées pour la sauvegarde des droits et libertés et des intérêts légitimes de la personne concernée, au moins du droit de la personne concernée d'obtenir une intervention humaine de la part du responsable du traitement, d'exprimer son point de vue et de contester la décision »[141]. Ce qui veut dire que dans certains cas, la personne concernée peut demander à un humain de revoir la décision automatisée. Une banque pourrait par exemple faire en sorte que le client puisse l'appeler pour demander un contrôle humain de la décision si la banque lui refuse un crédit par décision automatisée sur son site internet.

---

132 Korff 2012. Le prédécesseur était l'article 15 de la Directive sur la protection des données. Cet article était fondé sur une disposition de la loi française de 1978 sur la protection des données. Voir Bygrave 2001.

133 Le GDPR définit ainsi le profilage à son article 4(4) : toute forme de traitement automatisé de données à caractère personnel consistant à utiliser ces données à caractère personnel pour évaluer certains aspects personnels relatifs à une personne physique, notamment pour analyser ou prédire des éléments concernant le rendement au travail, la situation économique, la santé, les préférences personnelles, les intérêts, la fiabilité, le comportement, la localisation ou les déplacements de cette personne physique.

134 Art. 22 GDPR.

135 De Hert et Gutwirth 2008 ; Korff 2012 ; Wachter, Mittelstadt, et Floridi 2017 ; Zuiderveen Borgesius 2015a.

136 Mendoza et Bygrave 2017.

137 Voir Groupe de travail (Working Group) « article 29 » 2018 (WP251), p. 21.

138 Considérant 71 GDPR. Pour d'autres exemples qui pourraient constituer des décisions automatisées « affectant une personne de façon significative de manière similaire », voir Groupe de travail (Working Group) « article 29 » 2018 (WP251), p. 22.

139 Groupe de travail (Working Group) « article 29 » 2018 (WP251). p. 22.

140 Groupe de travail (Working Group) « article 29 » 2018 (WP251), p. 22.

141 Article 22(3) GDPR. Kaminski observe que le texte du GDPR crée une version algorithmique de la procédure régulière : un droit à la possibilité d'être entendu. Kaminski 2018a, p. 8.

Au-delà des exigences générales de transparence, le GDPR prévoit des exigences de transparence spécifique pour les décisions automatisées :

> Le responsable du traitement fournit à la personne concernée les informations suivantes : […] « l'existence d'une prise de décision automatisée, y compris un profilage […] et, au moins en pareils cas, des informations utiles concernant la logique sous-jacente, ainsi que l'importance et les conséquences prévues de ce traitement pour la personne concernée ».[142]

Dans certains cas, l'organisation devra donc indiquer qu'elle recourt à des décisions d'IA et fournir une information suffisante sur la logique du processus.

Les spécialistes universitaires ont beaucoup travaillé sur la question de savoir si les règles du GDPR en matière de décisions automatisées créent un « droit à l'explication » des décisions à caractère individuel[143]. Le considérant 71 pousserait à conclure à son existence — ce qui pourrait être un droit utile à la protection de l'impartialité[144].

De nombreux universitaires doutent de l'efficacité de ce droit, observant par exemple que beaucoup de formes de décision automatisée sortent du champ d'application des règles du GDPR[145]. Ces dernières ne s'appliquent par exemple qu'aux décisions prises « sur le seul fondement d'un traitement automatisé ». Si un employé de banque refuse un crédit « sur le seul fondement » de la recommandation du système d'IA, la disposition ne s'applique pas dès lors que l'employé ne se contente pas de valider le rejet[146].

Il faudra observer les effets pratiques de ces dispositions du GDPR. On l'a vu, la disposition figurant dans le texte qui avait précédé le GDPR en ce qui concerne les décisions automatisées est restée lettre morte. Quoi qu'il en soit, l'attention qu'a suscitée la disposition du GDPR a contribué à nourrir un débat interdisciplinaire sur l'explication des décisions d'IA.

La Convention 108 modernisée semble plus généreuse envers les personnes physiques dans la formulation des droits d'explication. Contrairement à la disposition du GDPR, qui s'applique aux décisions ayant des effets significatifs et prises « sur le seul fondement » d'un traitement automatisé, la Convention 108 donne à la personne concernée le droit « d'obtenir, à sa demande, connaissance du raisonnement qui sous-tend le traitement de données, lorsque les résultats de ce traitement lui sont appliqués »[147]. Il restera à voir comment la formule « les résultats de ce traitement lui sont appliqués » sera concrètement interprétée dans les mises en œuvre nationales.

---

[142] Article 13(2)(f) et 14(2)(f) GDPR.

[143] Voir par exemple Edwards et Veale 2017 ; Goodman et Flaxman 2016 ; Kaminski 2018 ; Kaminski 2018a ; Malgieri G et Comandé 2017 ; Mendoza et Bygrave 2017 ; Selbst et Powles 2017 ; Wachter *et al.* 2017.

[144] Considérant 71 : un traitement de ce type devrait être assorti de garanties appropriées : « le droit […] d'obtenir une explication quant à la décision prise à l'issue de ce type d'évaluation ».

[145] Voir par exemple Edwards et Veale 2017 ; Wachter *et al.* 2017 ; Zuiderveen Borgesius 2015a, chapitre 9, section 6.

[146] Voir Groupe de travail (Working Group) « article 29 » 2018 (WP251). Le Groupe de travail dit qu'un contrôle humain superficiel ne suffit pas. Mais, comme l'observent Veale et Edwards 2018, on ne voit pas clairement comment les organisations peuvent s'assurer que l'apport humain n'a pas été superficiel dans une décision.

[147] Article 9(1)(c) Convention du Conseil de l'Europe sur la protection des données 2018 : « Toute personne a le droit : […] d'obtenir, à sa demande, connaissance du raisonnement qui sous-tend le traitement de données lorsque les résultats de ce traitement lui sont appliqués ». Voir Veale et Edwards 2018.

### *Mises en garde*

Plusieurs mises en garde s'imposent en ce qui concerne les possibilités qu'offre le droit relatif à la protection de données comme instrument de lutte contre la discrimination issue de l'IA. Tout d'abord, l'insuffisance de la surveillance du contrôle du respect et de la répression. Les autorités de protection des données ont des ressources limitées. Beaucoup d'entre elles n'ont pas le pouvoir d'imposer des sanctions lourdes (le GDPR leur a conféré de nouveaux pouvoirs au sein de l'UE). Bon nombre d'organisations ne prennent pas très au sérieux le respect de la législation sur la protection des données[148]. Il semblerait que la conformité ait progressé avec le GDPR, mais il est encore trop tôt pour se prononcer.

Deuxièmement, certains algorithmes sortent du champ d'application de la législation sur la protection des données. Cette dernière ne couvre que le traitement de données personnelles, mais pas les modèles prédictifs, parce qu'ils ne portent pas sur des personnes identifiables. Par exemple, un modèle prédictif qui dit que 80 % des habitants d'un certain quartier paient leurs factures en retard ne comporte pas de données personnelles puisqu'il ne se réfère pas à des individus (la législation sur la protection des données s'applique en revanche dès que le modèle prédictif est appliqué à un individu[149]).

Troisièmement, la législation sur la protection des données comporte de nombreuses normes ouvertes et abstraites, plutôt que des règles bien tranchées[150]. Elle y est contrainte parce que ses dispositions s'appliquent à des situations très diverses, dans le secteur public et privé. Cette approche « tous azimuts » présente de nombreux avantages. Les normes ouvertes n'ont par exemple pas besoin d'être adaptées à l'apparition de toute nouvelle technologie. Mais leur inconvénient est qu'elles peuvent être difficiles à appliquer[151].

Quatrièmement, le droit relatif à la protection des données contient des règles strictes concernant les « catégories particulières » de données (parfois appelées « données sensibles »), comme l'origine raciale ou l'état de santé[152]. Ces règles entravent la détection et la réduction de la discrimination. De nombreuses méthodes de lutte contre la discrimination provoquée par les systèmes d'IA font l'hypothèse implicite que les organisations détiennent ces données sensibles — même si bien des organisations pourraient ne pas le faire pour se conformer à la législation sur la protection des données. Il est difficile de concilier le respect du droit relatif à la protection des données et la collecte de données sensibles à des fins de lutte contre la discrimination[153].

Cinquièmement, même si le GDPR ou la Convention 108 peuvent valablement exiger l'explication des décisions d'IA, il est souvent difficile d'expliquer la logique de la décision que prend un système d'IA par analyse d'un gros volume de données[154]. Et on ne voit pas très bien dans certains cas l'utilité de cette explication pour la personne

---

148 Voir Zuiderveen Borgesius 2015a, chapitre 8, section 2.

149 Voir Zuiderveen Borgesius 2015a, chapitre 2 et chapitre 5. Sur les faiblesses de la législation relative à la protection des données dans le domaine des décisions d'IA, voir Wachter et Mittelstadt 2018.

150 Zuiderveen Borgesius 2015a, chapitre 9, section 1.

151 Sur les différents types de règles juridiques, voir chapitre VI, section 1.

152 Article 9 GDPR ; article 6 Convention du Conseil de l'Europe sur la protection des données 2018. Les règles strictes relatives à des catégories particulières de données visent en partie à lutter contre la discrimination. Sur les catégories particulières de données dans le contexte de l'IA, voir Malgieri et Comandé 2017a.

153 Goodman 2016 ; Ringelheim et De Schutter 2008 ; Ringelheim et De Schutter 2009 ; Veale et Binns 2017 ; Žliobaitė et Custers 2016. Des méthodes d'audit des systèmes d'IA préservant la confidentialité des données par cryptage commencent à apparaître. Voir Kilbertus *et al.* 2018.

154 Ananny et Crawford 2016 ; Burrell 2016 ; Binns *et al.* 2018 ; Edwards et Veale 2017 ; Hildebrand 2015 ; Kroll *et al.* 2016 ; 2018 ; Wachter, Mittelstadt et Russell 2017.

concernée, dès lors surtout que c'est à elle de comprendre la décision et son bien-fondé[155].

Cela dit, un gain de transparence et l'explication des décisions d'IA pourraient se révéler utiles. Cela fait plus de dix ans que les spécialistes demandent que soient développées des technologies d'amélioration effective de la transparence des décisions automatisées[156]. Ces technologies devraient améliorer la transparence des flux d'information par le feed-back et la sensibilisation, ce qui permettrait aux personnes et aux groupements de mieux comprendre comment l'information est recueillie, agrégée, analysée et exploitée dans la prise des décisions[157]. Des informaticiens explorent actuellement diverses formes d'IA explicable[158].

Il est en tout cas bien trop tôt pour se prononcer sur les effets de la Convention 108 modernisée et du GDPR. Il faudra encore procéder à des études juridiques sur la façon dont la législation sur la protection des données pourrait contribuer à la réduction des risques de discrimination[159]. L'emploi du droit relatif à la protection des données dans la lutte contre la discrimination n'a jamais encore été largement testé, même s'il offre indéniablement des outils de lutte contre la discrimination illicite.

### 3. AUTRES NORMES

D'autres domaines du droit pourraient aussi contribuer à garantir la justice des décisions d'IA, et peut-être à réduire les problèmes de discrimination. La législation sur la protection des consommateurs pourrait par exemple être invoquée pour protéger les consommateurs contre des formes manipulatrices de publicité fondées sur l'IA[160]. Le comportement discriminatoire d'une entreprise causant des problèmes plus graves si elle détient un monopole, la législation sur la concurrence pourrait jouer un rôle dans la protection des personnes[161]. Dans le secteur public, le droit administratif et le droit pénal pourraient servir à préserver l'impartialité des procédures[162]. La législation sur la liberté d'information pourrait permettre d'obtenir des informations sur les systèmes d'IA du secteur public[163]. Mais l'application de toutes ces normes juridiques à la protection des personnes contre l'IA est encore largement inexplorée. Leur examen dépasserait le cadre du présent rapport.

#### *Réglementations envisagées*

Plusieurs réglementations susceptibles d'avoir des effets sur la discrimination suscitée par l'IA sont actuellement à l'étude. Le Comité consultatif de la Convention du Conseil de l'Europe pour la protection des personnes à l'égard du traitement automatisé des données à caractère personnel a publié en septembre 2018 un projet de rapport sur les problèmes suscités par l'intelligence artificielle dans la protection des données, et les solutions envisageables (*Artificial intelligence and data protection: challenges and possible remedies*[164]).

---

155 Edwards et Veale 2017.

156 Hildebrandt et Gutwirth 2008, chapitre 17.

157 Diaz et Gürses 2012.

158 Voir Guidotti *et al.* 2018 ; Miller 2017 ; Selbst et Barocas 2018 ; Tickle *et al.* 1998. Voir également le projet Google *What If... you could inspect a machine learning model, with no coding required ?*, https://pair-code.github.io/what-if-tool/index.html#about consulté le 1er octobre 2018. Ce projet s'est inspiré de Wachter, Mittelstadt et Russell 2017.

159 Des chercheurs commencent à étudier comment la législation relative à la protection des données peut contribuer à la lutte contre la discrimination. Voir par exemple : Goodman 2016 ; Mantelero 2018 ; Hacker 2018 ; Hoboken et Kostic (forthcoming) ; Wachter 2018 ; Wachter et Mittelstadt 2018.

160 Sur l'IA et le droit relatif à la protection des consommateurs, voir Contrôleur européen de la protection des données 2014 ; Helberger, Zuiderveen Borgesius et Reyna 2017 ; Jabłonowska *et al.* 2018.

161 Sur l'IA et le droit de la concurrence, voir Ezrachi et Stucke 2016 ; Graef 2016 ; Graef 2017 ; Valcke, Graef et Clifford 2018 ; Van Nooren *et al.* 2018.

162 Sur l'IA et le droit administratif, voir Van Eck 2018 ; Oswald 2018, Cobbe 2018.

163 Voir Rieke, Bogen et Robinson 2018, p. 24 ; Fink 2018 ; Oswald et Grace 2016.

164 Mantelero 2018.

Le Comité directeur sur les médias et la société de l'information du Conseil de l'Europe a formé le Comité d'experts sur la dimension droits de l'homme des traitements automatisés de données et différentes formes d'intelligence artificielle, qui procédera à des études et fournira des conseils en vue de la préparation éventuelle de normes pour l'avenir[165].

L'Union européenne se préoccupe aussi de l'IA. En 2018, la Commission européenne a publié une communication à ce sujet, et formé un groupe de haut niveau sur l'intelligence artificielle[166], chargé de préparer un projet de code de bonne conduite en matière d'IA[167]. L'Agence des droits fondamentaux de l'UE se penche également sur l'IA[168]. Et en 2017, la Commission a proposé un règlement sur la vie privée et les communications électroniques visant à protéger la vie privée sur l'internet ; le document pourrait affecter l'IA et l'apprentissage automatique en limitant la collecte de certains types de données sensibles sur l'internet[169].

Un règlement de 2016 de l'UE porte sur un type particulier de décision d'IA : le trading algorithmique en bourse, etc. Il veut que « les entreprises d'investissement s'assurent que les membres de leur personnel chargés de la vérification de la conformité comprennent au moins de façon générale la manière dont leurs systèmes de trading algorithmique et leurs algorithmes de négociation fonctionnent »[170], et que « les entreprises d'investissement appliquent un dispositif de gouvernance clair et formalisé, intégré à leur cadre global de gouvernance et de prise de décision »[171]. Il serait peut-être possible d'adopter des exigences comparables dans d'autres secteurs.

### *Autorégulation*

Plusieurs organisations ont proposé des principes visant à rendre l'IA équitable, transparente et éthique. La FATLM *(Fairness, Accountability, and Transparency in Machine Learning)* a par exemple publié des principes pour des algorithmes transparents et la déclaration de l'impact social des algorithmes[172]. Ces principes demandent aux organisations de veiller à ce que les décisions algorithmiques n'aient pas d'effet discriminatoire ou injuste sur les divers groupes sociaux (en fonction de la race, du sexe, etc.)[173].

Il existe d'autres principes d'autorégulation sur la déontologie et l'IA, souvent moins centrés sur la discrimination. Il y aurait par exemple les principes Asilomar AI du Future Life Institute (États-Unis)[174], la Déclaration de Montréal pour un développement responsable de l'IA[175], et les Principes pour une intelligence artificielle éthique de l'UNI Global Union[176]. L'IEE, une organisation professionnelle technique, a lancé une initiative mondiale sur la déontologie des systèmes autonomes et intelligents[177]. Et Apple, Amazon, DeepMind et Google, Facebook, IBM et Microsoft ont créé leur

---

165 Conseil de l'Europe MSI-AUT 2018.

166 https://ec.europa.eu/digital-single-market/en/high-level-expert-group-artificial-intelligence, consulté le 26 septembre 2018.

167 Commission européenne, L'intelligence artificielle pour l'Europe, 2018, p. 16. Voir également : European Group on Ethics in Science and New Technologies 2018.

168 http://fra.europa.eu/en/project/2018/artificial-intelligence-big-data-and-fundamental-rights, consulté le 13 octobre 2018.

169 Voir Zuiderveen Borgesius et al. 2017.

170 Règlement délégué (UE) 2017/589 de la Commission du 19 juillet 2016 complétant la directive 2014/65/UE du Parlement européen et du Conseil par des normes techniques de réglementation précisant les exigences organisationnelles applicables aux entreprises d'investissement recourant au trading algorithmique, paragraphe 2.

171 Article 1, *ibidem*.

172 https://www.fatml.org/resources/principles-for-accountable-algorithms, consulté le 24 septembre 2018.

173 https://www.fatml.org/resources/principles-for-accountable-algorithms, consulté le 24 septembre 2018.

174 https://futureoflife.org/ai-principles/ consulté le 24 septembre 2018.

175 https://www.montrealdeclaration-responsibleai.com/the-declaration, consulté le 24 septembre 2018.

176 http://www.thefutureworldofwork.org/opinions/10-principles-for-ethical-ai/, consulté le 24 septembre 2018.

177 https://standards.ieee.org/industry-connections/ec/autonomous-systems.html, consulté le 24 septembre 2018. Voir également Koene *et al.* 2018.

*Partnership on AI to Benefit People and Society* (partenariat pour mettre l'IA au service des personnes et de la société) pour étudier et formuler de bonnes pratiques en matière d'IA[178]. On peut en principe se féliciter de ces efforts d'autorégulation. Une IA éthique vaut mieux que le contraire. Les principes d'autorégulation pourraient contribuer à la réduction des problèmes de discrimination, et inspirer le législateur.

Mais la protection des droits de l'homme ne peut pas être laissée à l'autorégulation et à des normes non contraignantes[179]. Le problème majeur est aussi que les principes évoqués ci-dessus sont souvent assez abstraits, et ne contiennent pas de directives détaillées[180]. Wagner se méfie du « lavage éthique » de l'IA : le débat déontologique semble se centrer de plus en plus sur les façons qu'auraient les entreprises privées d'éviter l'intervention du législateur. Devant l'impossibilité ou le refus de la mise en place d'authentiques solutions réglementaires, l'éthique constitue la solution aisée pour structurer en douceur des actions d'autorégulation existantes et leur donner un sens[181]. Mais l'autorégulation et les normes non contraignantes ne doivent pas masquer l'éventuel besoin d'une vraie réglementation. Le chapitre VI examinera comment le droit pourrait être amélioré dans ce domaine. Nous formulons auparavant des recommandations à l'intention des organisations qui recourent à l'IA, des organes de contrôle du respect des droits de l'homme et des organismes de promotion de l'égalité.

## V. RECOMMANDATIONS

***Quelles mesures de réduction des risques de discrimination par l'IA peuvent être recommandées aux organisations qui l'utilisent, aux organismes de promotion de l'égalité des États membres du Conseil de l'Europe, et aux organes chargés de la surveillance du respect des droits de l'homme, comme la Commission européenne contre le racisme et l'intolérance ?***

### 1. ORGANISATIONS RECOURANT A L'IA

Il y a, pour les organisations publiques et privées désireuses de prévenir la discrimination, des mesures importantes à prendre lorsqu'elles recourent à l'IA : formation, obtention de compétences techniques et juridiques et préparation soigneuse des projets d'IA.

#### *Formation*

La formation joue un rôle important en sensibilisant l'organisation aux risques de discrimination accidentelle de l'IA. Le personnel concerné (cadres, juristes et informaticiens) doit être conscient de ces risques. Nous avons vu de nombreux exemples de discrimination par l'IA dans lesquels l'organisation n'avait pas d'intention discriminatoire. Si elle a conscience du risque, elle pourra prévenir la discrimination. Et la formation pourra peut-être aussi aider à combattre l'effet de « biais d'automatisation » au sein de son personnel[182].

#### *Évaluation et réduction des risques*

Au lancement d'un projet d'IA, l'organisation devrait procéder à l'évaluation et à la réduction des risques : a) en associant des spécialistes de plusieurs disciplines (informatique et droit, par exemple) à la détermination des risques du projet ; b) en enregistrant les évaluations et les mesures de réduction des risques ; c) en surveillant la mise en œuvre du projet ; d) en communiquant souvent l'information à l'extérieur (grand public ou organisme de surveillance)[183].

---

[178] https://www.partnershiponai.org/about/, consulté le 24 septembre 2018. Pour une liste et une critique d'autres principes éthiques pour l'IA, voir Greene, Hoffman, et Stark 2018.

[179] Pour une vue générale de l'autorégulation et des droits fondamentaux, voir Angelopoulos *et al.* 2016.

[180] Voir Campolo *et al.* 2017, p. 34.

[181] Wagner 2018. Voir également Nemitz 2018.

[182] Citron 2007, p. 1306. Sur le biais d'automatisation, voir Parasuraman et Manzey 2010 ; Rieke, Bogen et Robinson 2018, p. 11.

[183] Voir par exemple AI Now Institute 2018 ; Mantelero 2018 ; Mantelero 2018a.

L'organisation devrait obtenir l'appui d'informaticiens conscients des risques de discrimination (« informaticien » étant à comprendre ici au sens large : des spécialistes des données ou autres et des personnes suffisamment familiarisées avec l'IA pourraient aussi fournir leurs compétences). Les risques de discrimination dans les décisions d'IA commencent à former un nouveau domaine spécifique des sciences informatiques. Une organisation (la FATML) organise depuis 2014 des ateliers et des conférences pour rassembler un collectif grandissant de chercheurs et de praticiens soucieux de promouvoir l'équité, la redevabilité et la transparence dans l'apprentissage automatique[184]. Des spécialistes ont publié des résultats prometteurs, par exemple sur la prévention de la discrimination dans l'exploration de gros volumes de données[185].

Il peut être difficile de déterminer les risques d'un projet d'IA. Laissés à eux-mêmes, les informaticiens qui créent le système doivent prendre des décisions subjectives, et ont souvent du mal à faire comprendre les risques et les choix à l'équipe de direction[186]. Les personnes chargées du développement d'un système d'IA doivent être activement soutenues dans l'évaluation et la réduction des risques de discrimination ; le temps et les crédits nécessaires devraient être dûment pris en compte dans tout projet de ce type.

Les risques et les principes juridiques et normatifs applicables diffèrent d'un secteur à l'autre. Les risques ne sont par exemple pas les mêmes pour un système d'IA servant à sélectionner des candidats et pour un système de prédiction de la criminalité. Ce qui veut dire qu'il faudrait s'assurer le concours d'experts du secteur concerné[187]. Il peut être utile de créer un comité d'éthique qui évaluera et analysera les systèmes d'IA risquant de porter atteinte à des droits de l'homme[188]. Il peut également être utile d'aborder avec des universitaires, des groupes de la société civile et des personnes susceptibles d'être affectées les inquiétudes que leur inspire le système[189].

Une façon d'apprécier les risques d'un projet d'IA est de procéder à une forme appropriée d'analyse d'impact. On pourra s'inspirer de l'analyse d'impact exigée par le GDPR pour certains traitements de données à risque[190]. Et les organisations, surtout publiques, devraient envisager de publier le rapport de l'analyse d'impact.

Les risques présentés par le système d'IA devraient aussi être surveillés en cours d'utilisation, surtout que les phénomènes que reproduit le système changeront vraisemblablement au fil du temps, ce qui veut dire que les risques et l'impact changeront aussi[191]. Les organisations devraient envisager de publier des rapports annuels de surveillance du système.

Il est souvent possible de prévenir, ou du moins de minimiser, les effets discriminatoires. Une organisation peut par exemple décider de ne pas utiliser certaines caractéristiques comme données de base dans le système d'IA. À titre d'illustration, une société américaine d'aide au recrutement indique qu'elle n'utilise pas la distance domicile-travail dans la prédiction des candidats qui se révéleront de bons employés, car la corrélation avec la race est trop étroite. The Atlantic précise que la distance domicile-travail, par exemple, n'est jamais prise en compte dans le score d'un candidat, même si elle est communiquée à certains clients ; en effet, les quartiers et les villes peuvent présenter un certain profil racial, ce qui veut dire que la prise en compte de la distance pourrait enfreindre les normes d'égalité des chances au travail[192].

---

[184] https://www.fatml.org consulté le 1er août 2018.

[185] Voir par exemple Custers *et al.* 2013 ; Kamiran et Calders 2009 ; Kamiran et Calders 2012 ; Kusner *et al.* 2017 ; Pedreschi, Ruggieri, et Turini 2008.

[186] Veale, Van Kleek et Binns 2018. Kaminski 2018a, p. 30 : les ingénieurs ne devraient pas définir la discrimination ou l'équité sans une discussion approfondie avec des juristes et des membres de la collectivité.

[187] Campolo *et al.* 2017, p. 2.

[188] Conseil de l'Europe, Lignes directrices sur la protection des personnes à l'égard du traitement des données à caractère personnel à l'ère des mégadonnées, 2017, paragraphe 1.3.

[189] Voir article 35(9) GDPR.

[190] Conseil de l'Europe, Lignes directrices sur la protection des personnes à l'égard du traitement des données à caractère personnel à l'ère des mégadonnées, 2017, article 2.5. Voir également Reisman *et al.* 2018 ; Selbst 2017.

[191] Voir également article 35(110 GDPR ; Gama *et al.* 2014.

Le personnel des sociétés d'IA et des laboratoires de recherche universitaires est rarement très diversifié (le plus souvent masculin et blanc, par exemple). Une organisation de ce type pourrait accorder plus d'attention à la discrimination si son personnel était plus diversifié. Les organisations devraient donc veiller à employer un personnel plus diversifié[193]. Bien entendu, la diversité du personnel est toujours un objectif important.

***Organismes publics***

Le secteur public a plus de responsabilités que le privé. De nombreuses règles juridiques visent en effet à protéger la population contre le pouvoir de l'État, par exemple dans le domaine des droits de l'homme, des procédures pénales et du droit administratif. Ces responsabilités additionnelles s'appliquent aussi aux organismes publics qui utilisent des systèmes d'IA.

C'est pourquoi les systèmes d'IA du secteur public devraient être conçus dans un souci de transparence[194]. Parfois, l'information sur le système d'IA pourrait être communiquée au public pour contrôle, dans l'esprit du mouvement d'ouverture des données. Mais parfois aussi, cette information pourrait laisser transparaître des données personnelles et menacer le secret de la vie privée[195], ou permettre à certaines personnes de manipuler le système d'IA[196]. Les organismes publics pourraient donc ouvrir à des chercheurs ou à la société civile un accès contrôlé aux systèmes d'IA, dans un environnement sécurisé, comme les bureaux de statistiques le font aujourd'hui pour les microdonnées sensibles[197].

Le secteur public pourrait aussi adopter une clause de suppression automatique à la mise en place d'un système d'IA appelé à prendre des décisions affectant des personnes. Cette clause imposerait que le système soit soumis à une évaluation, par exemple au bout de trois ans, pour déterminer s'il a livré les résultats attendus[198]. Si ce n'est pas le cas, ou si les inconvénients ou les risques l'emportent, il faudrait envisager son abandon. Certes, le secteur public a plus de responsabilités, mais les organisations privées (entreprises, par exemple) pourraient prendre les mêmes mesures que celles qui ont été proposées pour le secteur public.

## 2. ORGANISMES DE PROMOTION DE L'EGALITE ET ORGANES DE SURVEILLANCE DU RESPECT DES DROITS DE L'HOMME

***Quelles mesures de réduction des risques de discrimination par l'IA peuvent être recommandées aux organismes de promotion de l'égalité des États membres du Conseil de l'Europe, et aux organes chargés de la surveillance du respect des droits de l'homme, comme la Commission européenne contre le racisme et l'intolérance ?***

***Formation et compétences techniques***

Les organismes de promotion de l'égalité et les organes chargés de la surveillance du respect des droits de l'homme devraient comprendre les promesses et les menaces de l'IA. Il est donc nécessaire de leur fournir une formation de base sur l'IA et ses risques.

---

192 Peck 2013. Voir également Rieke, Robinson et Yu 2014, p. 15.

193 Campolo *et al.* 2017, p. 16.

194 Voir Kroll *et al.* 2016 ; Munoz, Smith et Patil, 2016.

195 Veale, Binns et Edwards 2018.

196 Laskov et Lippman 2011 ; Bambauer et Zarsky 2018.

197 Sur les degrés variables de liberté dans le contexte des données libres, voir Zuiderveen Borgesius, Gray, et Van Eechoud 2015.

198 McCray, Oye et Petersen 2010 ; Broeders, Schrijvers et Hirsch Ballin, p. 23.

Les organismes de promotion de l'égalité et les organes de surveillance du respect des droits de l'homme devraient aussi veiller à disposer des compétences techniques nécessaires en matière d'IA en s'assurant le concours de spécialistes des sciences informatiques[199]. Ces derniers sont mieux à même de percevoir et de comprendre certains risques que, par exemple, des juristes[200]. Même si ce ne sont pas des spécialistes de l'IA, ils pourraient procéder eux-mêmes à certaines recherches sur la discrimination provoquée par l'IA. Comme l'observent Rieke, Bogen et Robinson, la surveillance n'a pas besoin d'être très complexe pour donner de bons résultats[201]. Les problèmes que présente un système d'IA peuvent souvent être découverts par la simple observation de ses intrants et extrants[202]. Et les informaticiens non spécialisés dans l'IA savent souvent à quels spécialistes faire appel dans une enquête. Selon leur budget, les organismes de promotion de l'égalité et les organes de surveillance du respect des droits de l'homme pourraient recruter des informaticiens pour un projet, ou de façon plus permanente.

Ils pourraient envisager d'organiser des campagnes de sensibilisation des organisations publiques et privées[203]. On l'a vu, c'est souvent par accident que les organisations utilisent des systèmes d'IA discriminatoires. La sensibilisation pourrait se révéler utile.

D'une façon plus générale, les écoles et les universités qui enseignent l'informatique, la science des données, l'IA et des matières connexes devraient familiariser aussi leurs étudiants avec les droits de l'homme et la déontologie. Beaucoup d'universités offrent déjà des cours de ce type à leurs étudiants d'informatique[204]. Les organismes de promotion de l'égalité et les organes de surveillance du respect des droits de l'homme pourraient envisager d'aider les écoles et les universités à organiser ces cours[205].

Il serait possible de favoriser le débat public en familiarisant le grand public avec les risques de discrimination par l'IA, mais aussi avec les nombreux avantages et possibilités qu'elle offre. La sensibilisation ne devrait cependant pas conduire à la responsabilisation, c'est-à-dire à la responsabilité individuelle d'une mission précédemment confiée à une autre entité (normalement de l'État) ou simplement non reconnue auparavant comme une responsabilité[206]. On ne saurait remettre à l'individu le soin de se défendre lui-même contre la discrimination[207]. La sensibilisation n'en est pas moins nécessaire à un débat inclusif sur les risques des décisions d'IA.

### *Consultation préalable des organismes de promotion de l'égalité*

Les organismes de promotion de l'égalité pourraient demander aux organes publics de leur soumettre leurs projets de décisions automatisées affectant des personnes ou des groupes. Un tel organisme pourrait par exemple les aider à vérifier si les données d'apprentissage sont biaisées[208]. Il pourrait aussi demander à chaque organe public utilisant une IA pour des décisions automatisées affectant des personnes de s'assurer qu'il possède les compétences juridiques et techniques nécessaires à l'évaluation et la surveillance du risque. Et il pourrait aussi être exigé des organismes publics qu'ils vérifient régulièrement que leur système d'IA ne produit pas d'effets discriminatoires. (Selon le contexte national, les organismes de promotion de l'égalité pourraient aussi suggérer plutôt qu'exiger.)

---

[199] Comme nous l'avons dit, l'expression « spécialistes des sciences informatiques » est ici une étiquette générale : des spécialistes des données et d'autres personnes possédant une connaissance suffisante de l'IA pourraient également apporter leurs compétences.

[200] Sur l'importance des compétences techniques des autorités de protection des données, voir Raab et Szekely 2017.

[201] Rieke, Bogen et Robinson 2018, p. 2.

[202] Rieke, Bogen et Robinson 2018, p. 8. On y trouvera également des exemples de surveillance de systèmes d'IA (p. 31-34).

[203] Voir Résolution de 2002 relative au statut de l'ECRI, article 12 ; ECRI, Recommandation de politique générale n° 2 (2018), paragraphe 13(e) ; paragraphe 34, et exposé des motifs, paragraphe 64.

[204] Fiesler 2018 a dressé une liste de plus de 200 cours de déontologie technique.

[205] Voir ECRI, Recommandation de politique générale n° 10 sur la lutte contre le racisme et la discrimination raciale dans et travers l'éducation scolaire, 15 décembre 2006, Strasbourg, CRI(2007)6 https://rm.coe.int/ecri-general-policy-recommendation-no-10-on-combating-racism-and-racia/16808b5ad5, consulté le 14 octobre 2018.

[206] Wakefield et Fleming 2009.

[207] Voir également Ellis et Watson 2012, p. 502-503.

[208] Voir ECRI Recommandation de politique générale n° 2 (2018), paragraphe 13(g). Voir également article 36 GDPR sur la consultation préalable.

Les organismes de promotion de l'égalité et les organes de surveillance du respect des droits de l'homme pourraient contribuer à la conception d'une méthode spécifique d'évaluation des droits de l'homme et de l'impact de l'IA. Comme indiqué précédemment, l'analyse d'impact peut être utile, mais il n'en existe pas pour l'instant de forme spécifiquement adaptée à l'IA[209]. Diverses parties prenantes et spécialistes de plusieurs disciplines devraient être associés à sa conception. Il serait possible de s'inspirer des analyses d'impact sur la vie privée et la protection des données[210].

### *Participation à la passation des marchés publics*

Les organismes de promotion de l'égalité devraient s'appuyer sur les dispositions et processus nationaux et faire pression pour obtenir un meilleur accès afin de se faire associer très précocement à la passation de marchés publics d'acquisition de systèmes d'IA. Ils peuvent veiller à ce que la conception de ces systèmes tienne compte des inquiétudes suscitées par les risques de discrimination, et à ce que les systèmes soient suffisamment ouverts aux audits et fassent l'objet de garanties appropriées.

### *Coopération avec les autorités de protection des données*

Nous avons dit que les deux instruments juridiques les plus utiles, en ce qui concerne la discrimination provoquée par l'IA, sont la législation anti-discrimination et la législation sur la protection des données. Il serait extrêmement dommage de ne pas les combiner[211]. Les organismes de promotion de l'égalité devraient coopérer avec les autorités de protection des données. Il pourrait être utile, par exemple, qu'ils échangent leurs savoirs et leur expérience[212]. Nombre d'autorités de protection des données possèdent certaines compétences techniques au sein de leur personnel[213], et certaines ont l'expérience du recrutement d'informaticiens extérieurs pour des projets de recherche[214]. Les autorités de protection des données pourraient par exemple, si elles ont vent d'organisations qui utilisent des systèmes d'IA présentant un risque de discrimination, alerter l'organisme de promotion de l'égalité. Ce dernier pourrait communiquer des informations aux autorités de protection des données, par exemple sur les risques de discrimination. Selon la situation constatée dans chaque pays, il pourrait aussi être utile que des organismes de promotion de l'égalité coopèrent avec les autorités de protection des consommateurs et les autorités de surveillance de la concurrence.

Au chapitre de la coopération et du partage de savoirs entre les organes de régulation, le Contrôleur européen de la protection des données a proposé en 2016 de créer un réseau volontaire d'organismes de contrôle leur permettant de partager des informations sur les abus possibles de l'écosystème numérique et les formes les plus efficaces de lutte[215]. Cette initiative pourrait servir d'inspiration aux organismes de promotion de l'égalité et aux organes de surveillance du respect des droits de l'homme[216].

---

[209] Reisman *et al.* 2018 examinent l'évaluation d'impact algorithmique aux États-Unis. Les États membres du Conseil de l'Europe ayant des systèmes juridiques différents de celui des États-Unis, ces méthodes pourraient les inspirer, mais ne seraient pas directement transférables.

[210] Voir Binns 2017 ; Kloza *et al.* 2017 ; Mantelero 2017 ; Wright et De Hert 2012. Voir également the Brussels Laboratory for Data Protection and Privacy Impact Assessments http://dpialab.org/.

[211] Voir Schreurs *et al.* 2008 ; Gellert *et al.* 2013 ; Hacker 2018 ; Lammerant, De Hert, Blok 2017.

[212] Voir ECRI, Recommandation de politique générale no 2 (2018), paragraphe 13(b).

[213] Raab et Szekely 2017.

[214] Des chercheurs de l'université de Louvain ont par exemple examiné le système de suivi de Facebook pour l'Autorité belge de protection des données. Voir Autorité belge de protection des données 2018.

### *Coopération avec le monde universitaire*

Les organismes de promotion de l'égalité et les organes de surveillance du respect des droits de l'homme devraient maintenir des contacts, voire coopérer, avec les universitaires. Notre rapport révèle de nombreux exemples de décisions discriminatoires d'AI découvertes par des chercheurs universitaires (et des journalistes d'investigation)[217]. Les universitaires sont souvent heureux d'aider les organes de régulation, mais n'ont pas de contacts réguliers avec eux. Dans un premier temps, les organismes de promotion de l'égalité et les organes de surveillance pourraient assister à des conférences et autres rencontres de chercheurs universitaires. La discrimination par l'IA est très souvent abordée dans des conférences internationales sur la protection de la vie privée. Plusieurs de ces conférences attirent des organismes de régulation, des praticiens, des groupes de la société civile et des universitaires de différentes disciplines (droit, informatique, philosophie et sociologie)[218]. Les organismes de promotion de l'égalité et les organes de surveillance pourraient aussi envisager d'organiser des conférences, des tables rondes et d'autres rencontres sur les risques de discrimination par l'IA, pour encourager les contacts entre les chercheurs et les organes de promotion de l'égalité. Et peut-être est-ce que ces derniers et les organes de surveillance pourraient faire procéder à davantage de recherches sur les risques de discrimination par l'IA (se reporter à ce sujet à la section VI.3), ou former un groupe de travail à ce sujet[219].

Les organismes de promotion de l'égalité et les organes de surveillance du respect des droits de l'homme devraient se rapprocher non seulement des groupes de la société civile travaillant sur la discrimination[220], mais aussi des associations de consommateurs[221] et des groupes de la société civile spécialisés dans les politiques technologiques et les droits numériques[222]. Les groupes de la société civile qui travaillent sur la discrimination possèdent souvent des compétences différentes de ceux qui travaillent sur la technologie et les droits numériques. Il serait utile de multiplier les contacts entre eux, car beaucoup s'intéressent aussi à la discrimination causée par l'IA[223].

---

[215] Contrôleur européen de la protection des données 2016.

[216] D'ailleurs, il faudrait aussi intensifier la coopération entre juristes de différentes disciplines au sein des universités : spécialistes du droit relatif à la non-discrimination (qui travaillent fréquemment dans des instituts des droits de l'homme) et du droit relatif à la protection des données (souvent rattachés à des instituts de droit et de technologie).

[217] Voir Rieke, Bogen et Robinson 2018, p. 31.

[218] Voir par exemple CPDP Computers, Privacy and Data Protection conference in Brussels https://www.cpdpconferences.org ; APC Amsterdam Privacy Conference https://www.apc2018.com ; TILTing Perspectives https://www.tilburguniversity.edu/research/institutes-and-research-groups/tilt/events/tilting-perspectives ; et PLSC Privacy Law Scholars Conference http://law.berkeley.edu/plsc. L'ACM Conference on Fairness, Accountability, and Transparency (ACM FAT*) se tiendra à Amsterdam en 2020 : https://www.fatml.org . Consultés le 14 octobre 2018.

[219] Voir Résolution de 2002 relative au statut de l'ECRI, article 6(1) et 6(2) ; ECRI, Recommandation de politique générale no 2 (2018), paragraphe 13(d).

[220] Voir Résolution de 2002 relative au statut de l'ECRI, articles 10(1) et 13.

[221] Le BEUC (Bureau européen des unions de consommateurs) pourrait constituer un point de contact avec les associations de consommateurs. Il rassemble 43 associations de consommateurs de 32 pays européens. HTTPS://www.beuc.eu/about-beuc/who-we-are, consulté le 10 octobre 2018. Voir également Bureau européen des unions de consommateurs BEUC 2018.

[222] En ce qui concerne les groupes centrés sur les droits et libertés dans l'environnement numérique, un point de contact pourrait être Droit numérique européen (EDRi), une association d'organisations de défense des droits civils et des droits de l'homme de toute l'Europe. https://edri.org/members/, consulté le 10 octobre 2018.

[223] Voir Gangadharan et Niklas 2018, qui ont interrogé des ONG et conclu à la nécessité d'une meilleure coopération entre les ONG spécialisées dans la vie privée et la technologie d'une part, et les ONG spécialisées dans la discrimination de l'autre.

***Traitement des contentieux et réglementation***

En fonction de la situation nationale, les organismes de promotion de l'égalité pourraient aussi assurer le traitement stratégique des contentieux dans le domaine des décisions d'IA[224]. Avec les organismes de surveillance du respect des droits de l'homme, ils pourraient demander l'adoption de normes réduisant les risques de discrimination par l'IA[225]. Le chapitre suivant propose des améliorations possibles en la matière.

# VI. AMÉLIORATION DE LA RÉGLEMENTATION

***Par quels dispositifs (juridiques, réglementaires, d'autorégulation) est-il possible de réduire les risques ?***

On a vu au chapitre IV que la législation actuelle présente des points faibles en matière de discrimination provoquée par l'IA. D'autres textes seront probablement nécessaires pour protéger les personnes contre la discrimination illicite et la différenciation injuste. La section 1 présente des observations liminaires sur la réglementation d'une technologie en rapide développement. La section 2 se penche sur l'amélioration du respect des normes actuelles de lutte contre la discrimination. La section 3 examine si les décisions d'IA appellent une modification des normes juridiques elles-mêmes. Les idées proposées dans ce chapitre sont à considérer comme des pistes de réflexion, et non comme des conseils à proprement parler pour la définition de politiques en la matière.

## 1. LA REGLEMENTATION DE TECHNOLOGIES EN RAPIDE ESSOR

La réglementation d'une technologie en rapide essor pose des difficultés particulières. L'adoption de textes juridiques ou de traités peut prendre des années, voire des dizaines d'années. Pendant ce temps, la technologie concernée, le marché et la société évoluent rapidement.

Ces difficultés ne concernent pas spécifiquement l'IA, et il existe une certaine expérience de la réglementation de technologies nouvelles. Dans ce domaine, les politiques peuvent combiner des normes variées : textes législatifs pour poser les grands principes, et lignes directrices (émanant des organes de régulation, par exemple) fixant des règles plus précises[226]. La loi pourrait être formulée d'une façon raisonnablement indépendante de la technologie. Les dispositions de ce type définissent les grands principes et ont pour avantage de ne pas nécessiter de modification à chaque avancée de la technologie ; en revanche, elles peuvent se révéler difficiles d'application dans la pratique. Les lignes directrices formulées par l'organe de régulation peuvent donc se révéler utiles[227] ; elles peuvent être modifiées plus rapidement, ce qui leur permet d'être plus spécifiques et concrètes ; elles devraient faire l'objet d'évaluations régulières, et être modifiées dès que cela devient nécessaire[228].

La réglementation de la protection des données adopte en partie cette approche combinée[229]. Les textes juridiques (comme le GDPR et la Convention 108 modernisée) contiennent de nombreuses dispositions à la formulation large, applicables à diverses situations et technologies[230]. Ils ne prévoient par exemple pas de règles spécifiques pour la vidéosurveillance ou pour la surveillance sur le lieu de travail. Mais ils s'appliquent à ces deux cas dès lors qu'il y a données personnelles (notamment des images vidéo).

---

[224] Voir ECRI, Recommandation de politique générale nº 2 (2018), paragraphes 14-16.

[225] Voir Résolution de 2002 relative au statut de l'ECRI, article 1 ; ECRI, Recommandation de politique générale no 2 (2018), paragraphe 13(j).

[226] Voir Koops 2006.

[227] Voir Zuiderveen Borgesius 2015a, chapitre 9, section1 ; Baldwin, Cave et Lodge 2011, chapitre 14.

[228] Voir Koops 2006.

[229] Je ne dis pas que la législation sur la protection des données devrait être considérée comme une bonne pratique de réglementation des domaines dans lesquels une technologie se développe rapidement. La législation sur la protection des données est loin d'être irréprochable.

Outre les textes juridiques sur la protection des données, les autorités de protection des données adoptent fréquemment des lignes directrices interprétatives fixant des exigences plus spécifiques et plus concrètes pour des situations diverses, comme la vidéosurveillance[231], le lieu de travail[232] et les décisions automatisées[233]. Au sein de l'UE, le Comité européen de la protection des données et son prédécesseur ont adopté plus de 250 lignes directrices depuis 1995[234]. De même, le Conseil de l'Europe a adopté des lignes directrices en plus de la Convention 108 sur la protection des données, par exemple sur les mégadonnées (*Big Data)*[235], la police[236] et le profilage[237].

Ce qui veut dire que pour réduire par de nouvelles règles les risques de discrimination par l'IA, il conviendrait sans doute de combiner des normes juridiques à des possibilités offertes aux organes de régulation de préparer des lignes directrices plus aisées à modifier. Il existe d'autres possibilités que les textes de loi et les textes d'orientation des organes de contrôle, comme la corégulation : de l'autorégulation soumise à divers degrés d'influence des organes publics de contrôle. L'idée de base reste la même : combiner des règles de nature différente[238]. Comme le dit Koops, la réglementation multiniveaux, les formulations ouvertes et un mélange de règles abstraites et concrètes régulièrement réévaluées permettent d'obtenir une bonne sécurité juridique en ce qui concerne les technologies du moment, tout en laissant une marge suffisante de développement aux technologies concernées[239].

Évidemment, les entités chargées de formuler les règles et les lignes directrices doivent posséder une légitimité démocratique, avec un dispositif suffisant de freins et de contrepoids. La réglementation de nouvelles technologies est donc difficile, mais possible — et souvent nécessaire.

## 2. CONTROLE DU RESPECT

### *Meilleur respect des normes anti-discrimination existantes*

Les normes générales applicables à la discrimination causée par les décisions d'IA sont raisonnablement claires : notre société n'accepte pas ni ne doit accepter la discrimination au motif des caractéristiques protégées, comme l'origine raciale. Nous suggérons ci-dessous quelques moyens de faire respecter les normes anti-discrimination dans le domaine de l'IA.

---

[230] La législation sur la protection des données mise en place à partir du début des années 1970 pourrait elle-même être perçue comme la réponse juridique à un phénomène nouveau : les administrations de grande envergure et le traitement automatisé des données personnelles. Depuis ses débuts, la législation sur la protection des données a été continuellement adaptée au fil de l'évolution du domaine, comme en témoignent le récent GDPR et la Convention 108 modernisée.

[231] Groupe de travail (Working Group) « article 29 » 2004 (WP89).

[232] Groupe de travail (Working Group) « article 29 » 2017 (WP249).

[233] Groupe de travail (Working Group) « article 29 » 2018 (WP251).

[234] Voir site internet du Comité européen de la protection des données https://edpb.europa.eu/edpb_en. Son prédécesseur était appelé Groupe de travail (Working Group) « article 29 » http://ec.europa.eu/newsroom/article29/news.cfm?item_type=1308. Les avis et lignes directrices sont également réunis sur le site https://iapp.org/resources/article/all-of-the-european-data-protection-board-and-article-29-working-party-guidelines-opinions-and-documents/. Liens consultés le 10 octobre 2018.

[235] Conseil de l'Europe, lignes directrices sur les mégadonnées 2017.

[236] Conseil de l'Europe, Guide pratique sur l'utilisation de données à caractère personnel dans le secteur de la police, 2018.

[237] Conseil de l'Europe, Recommandation sur le profilage 2010.

[238] Voir Angelopoulos *et al.* 2016, p. 5-6 ; Brown et Marsden 2013. Et sur la corégulation : Hirsch 2010 ; Kaminsky 2018a.

[239] Koops 2006.

***Transparence***

Nous avons déjà mentionné le manque de transparence parmi les problèmes suscités par les systèmes d'IA, qui fonctionnent comme des « boîtes noires »[240]. Cette opacité est un problème en soi, mais elle freine aussi la découverte des effets discriminatoires.

La réglementation (ce qui englobe les lignes directrices, etc.) pourrait viser à améliorer la transparence, par exemple en exigeant que la conception des systèmes d'IA du secteur public permette les audits et les explications[241]. Ces exigences pourraient aussi s'appliquer aux acteurs privés[242]. Il existe des précédents en la matière dans le secteur privé : certains algorithmes de trading doivent être interprétables[243].

Pour certains types de systèmes, il pourrait être utile que les organes publics publient le code source du logiciel. Son examen peut parfois fournir des informations sur le fonctionnement du système. Rieke, Bogen et Robinson observent que les audits de code source ont les plus grandes chances d'être utiles lorsqu'il s'agit de répondre à une question claire sur le fonctionnement d'un logiciel dans un espace réglementé, avec des normes particulières sur la base desquelles juger le comportement ou les performances d'un système[244]. Il serait possible de modifier les lois sur la liberté de l'information de façon à leur faire couvrir le code source des systèmes d'IA. Cette modification permettrait aux journalistes, aux universitaires et à d'autres d'obtenir et d'examiner le code source.

Les systèmes d'IA sont protégés par le secret des affaires, les droits de propriété intellectuelle ou les conditions de vente des entreprises[245]. Cette protection entrave les organes de contrôle, les journalistes et les universitaires dans l'examen des systèmes. La législation pourrait être modifiée et admettre davantage d'exceptions pour motif de recherche, de façon à rendre possibles certains types de recherche. Et elle devrait peut-être aussi imposer aux organisations de fournir certaines informations aux chercheurs à leur demande. Les textes devront judicieusement arbitrer entre l'intérêt public de transparence et l'intérêt commercial, la confidentialité et autres intérêts protégés par l'opacité[246].

Bien souvent, le code source ne renseigne pas beaucoup à lui seul sur le système d'IA, dont l'évaluation n'est possible que par observation de son fonctionnement dans la réalité. Même pour des programmes modérément complexes, constatent Rieke, Bogen et Roginson, il peut se révéler nécessaire de voir fonctionner un programme dans des conditions réelles, avec de vrais utilisateurs et de vraies données, pour véritablement appréhender ses effets[247].

La législation pourrait imposer au secteur public de n'utiliser que des systèmes d'IA dont les risques ont été convenablement évalués, et qui permettent la surveillance et les audits[248]. On pourrait envisager une exigence analogue pour le secteur privé lorsque les systèmes d'IA servent à prendre certaines décisions (assurance, crédit ou emploi, par exemple)[249]. Des recherches et des débats sont encore nécessaires pour déterminer qui procéderait à ces audits. Une organisation chargée du contrôle et de l'audit de systèmes d'IA a besoin de compétences considérables[250].

---

240 Pasquale 2015. Voir également Zarsky 2018.

241 Voir Rieke, Bogen et Robinson 2018, p. 6 ; Pasquale 2017. Sur l'audit des systèmes d'IA, voir Sandvig *et al.* 2014.

242 Voir Rieke, Bogen et Robinson 2018, p. 6.

243 Voir section IV.3 ; partie sur le trading algorithmique.

244 Rieke, Bogen et Robinson 2018, p. 19.

245 Voir Bodo *et al.* 2017, p. 171-175 ; Malgieri 2016 ; Wachter et Mittelstadt 2018, p. 63-77.

246 Des questions similaires se posent en ce qui concerne les données libres et la protection de la confidentialité. Voir Zuiderveen Borgesius, Gray, et Van Eechoud 2015.

247 Rieke, Bogen et Robinson 2018, p. 19.

248 Voir Campolo *et al.* 2018, p. 1.

249 Voir Campolo *et al.* 2018, p. 1.

***Pouvoirs d'investigation et de répression***

Les États membres du Conseil de l'Europe devraient faire en sorte que les organismes de promotion de l'égalité et les autorités de protection des données soient convenablement financés et disposent de pouvoirs suffisants d'investigation et de répression[251]. Sans répression des infractions, la transparence ne conduira pas nécessairement à la responsabilisation[252].

En conclusion, les organismes de promotion de l'égalité et les organes de surveillance du respect des droits de l'homme peuvent demander que la réglementation impose mieux le respect des normes anti-discrimination actuelles dans le domaine des décisions d'IA. Mais ces dernières ouvrent sur de nouvelles formes de discrimination et de différenciation qui échappent largement à la réglementation actuelle, notamment de lutte contre la discrimination. C'est la question sur laquelle nous allons nous pencher à présent.

## 3. LA REGLEMENTATION DE NOUVELLES DIFFERENCIATIONS

Le dispositif de réglementation de la non-discrimination et de la protection des données comporte des lacunes en ce qui concerne l'IA[253]. De nombreux textes de lutte contre la discrimination ne couvrent que certaines caractéristiques protégées, comme la race, le genre ou l'orientation sexuelle[254]. Ils ne s'appliquent pas à la discrimination au motif de l'état de fortune, par exemple. La législation sur la protection des données pourrait aider à combler certaines de ces lacunes, mais certainement pas toutes.

Les systèmes d'IA peuvent échapper aux dispositions anti-discrimination lorsqu'ils différencient des catégories non encore répertoriées[255]. Pour prendre un exemple simple, supposons qu'un système d'IA découvre une corrélation entre l'utilisation d'un certain navigateur et l'acceptation de prix supérieurs. Un magasin en ligne pourrait demander des prix plus élevés aux utilisateurs de ce navigateur[256]. Cela ne relèverait pas de la législation anti-discrimination, le navigateur ne faisant pas partie des caractéristiques protégées (nous faisons ici l'hypothèse que le navigateur ne peut pas servir de donnée indirecte de détection d'une caractéristique protégée).

***L'IA peut renforcer les inégalités sociales***

Les décisions d'IA peuvent procéder à des différenciations injustes ou présentant d'autres effets indésirables même si elles n'entrent pas dans le champ d'application de la législation anti-discrimination. Une compagnie d'assurances, par exemple, pourrait utiliser un système d'IA pour fixer les primes de chaque assuré ou pour refuser une police. La différenciation par le risque est dans une certaine mesure nécessaire, et c'est une pratique acceptée chez les assureurs. On peut considérer comme équitable qu'un assuré présentant un risque supérieur verse une prime supérieure.

Mais il y a des inconvénients. La différenciation exagérée par le risque pourrait rendre l'assurance impossible à certains consommateurs et compromettre la fonction de mutualisation du risque de l'assurance. Elle pourrait aussi avoir pour effet de faire payer davantage aux pauvres. Un consommateur qui habite dans un quartier pauvre, où les cambriolages sont nombreux, paiera une prime d'assurance supérieure pour son logement, du fait que le risque de vol est plus élevé. Mais si les quartiers où habitent de nombreux pauvres présentent un risque supérieur, les pauvres paieront davantage en moyenne[257].

---

[250] Il a été suggéré qu'un organisme spécial de surveillance du profilage automatisé (décisions d'IA) pourrait être utile. Voir Koops 2008.

[251] Voir ECRI Recommandation de politique générale nº 2 (2018), paragraphe 28.

[252] Voir Kaminski 2018a, p. 21.

[253] Voir sections IV.1 et IV.2.

[254] Gerards 2007 ; Khaitan 2015.

[255] Custers 2004. Voir également Mittelstadt *et al.* 2016.

[256] Il n'y a pas d'indice de recours à cette pratique, mais cette discrimination par le prix serait techniquement aisée. Il a existé toutefois un site de voyages et de réservations qui présentait des hôtels plus chers aux utilisateurs de matériel Apple, et moins chers aux utilisateurs de PC (Mattioli 2012).

L'IA pourrait renforcer les inégalités sociales d'une façon plus générale. Valentino-De Vroes, Singer-Vine et Soltani ont par exemple montré que certaines pratiques de différenciation des prix en ligne aux États-Unis avaient pour effet que les habitants de zones d'habitation pauvres paient des prix plus élevés. Plusieurs magasins demandaient plus aux personnes vivant à la campagne qu'aux citadins[258]. À la campagne, le consommateur doit parcourir une longue distance pour se rendre chez un concurrent. Un magasin en ligne n'a donc pas besoin de recourir à des prix bas : la plupart des consommateurs ne feront pas le trajet pour obtenir un meilleur prix. Dans une grande ville, le consommateur peut aisément acheter le même produit chez un concurrent ; certains magasins en ligne y pratiquent donc des prix inférieurs. Ce mode de formation des prix avait pour effet — sans doute non intentionnel — que les gens pauvres payaient en moyenne davantage, du fait que les personnes habitant à la campagne ont tendance à être plus pauvres aux États-Unis[259]. L'IA peut donc renforcer les inégalités sociales. Or l'état de fortune ne constitue pas une caractéristique protégée, et la législation anti-discrimination ne couvre donc pas une telle pratique (à moins qu'elle ne débouche sur une discrimination indirecte fondée sur une caractéristique protégée)[260].

### *L'IA peut se tromper*

Le droit relatif à la discrimination n'aborde guère les erreurs de prédiction de l'IA (faux positifs ou négatifs). Or les décisions d'IA sont souvent erronées à l'échelon individuel, car elles appliquent fréquemment un modèle prédictif à un individu. Prenons un exemple : 80 % des personnes ayant le code postal XYZ règlent leurs factures en retard. Si, sur cette base, une banque refuse d'accorder un crédit à tous les habitants de ce quartier, son refus s'applique aussi aux 20 % d'habitants qui règlent leurs factures à temps[261]. Une telle pratique pourrait pénaliser de façon disproportionnée certains groupes sociaux. Un système d'AI fait parfois plus d'erreurs parmi les groupes minoritaires que dans la population majoritaire[262].

### *Nouvelles normes ?*

Il faudrait envisager de compléter la réglementation parce que les décisions d'IA non couvertes par les lois anti-discrimination peuvent être injustes. Mais il ne serait probablement pas très utile d'adopter des règles applicables à l'ensemble des décisions de ce type. L'IA est en effet utilisée dans bien des secteurs et à bien des fins, et menace rarement les droits de l'homme[263]. L'IA d'un jeu d'échecs ne comporte pas les mêmes risques qu'un système de prédiction policière par IA.

---

[257] Sur l'IA dans l'assurance, voir Dutch Association of Insurers 2016 ; Financial Conduct Authority 2016 ; Peppet 2014 ; Swedloff 2014. L'Allemagne a adopté des règles spécifiques sur les décisions automatisées dans l'assurance. Voir Bundesdatenschutzgesetz (loi fédérale sur la protection des données) du 20 juin 2017 (BGBl. I S. 2097), section 37 https://www.gesetze-im-internet.de/englisch_bdsg/englisch_bdsg.html#p0310 consulté le 13 octobre 2018. Voir également Malgieri 2018, p. 9-11. Sur la discrimination dans l'assurance, voir Avraham 2017.

[258] Valentino-De Vries, Singer-Vine et Soltani 2012.

[259] Valentino-De Vries, Singer-Vine et Soltani 2012. Sur le renforcement des inégalités sociales et le « tri social », voir également Atrey 2018 ; Danna et Gandy 2002 ; Lyon 2002 ; Naudts 2017 ; Taylor 2017 ; Turow 2011. Gandy mettait en garde il y a 25 ans déjà contre les effets discriminatoires du traitement de données à grande échelle (Gandy 1993).

[260] Le droit sur la protection des données ne traite pas l'état de fortune d'une personne comme une catégorie particulière de données (article 9 GDPR).

[261] Zarsky 2002.

[262] Pour un exemple de système commettant davantage d'erreurs dans les minorités, voir Rieke, Robinson et Yu 2014, p. 12. En pareil cas, la décision d'IA pourrait constituer une forme de discrimination indirecte interdite. Voir également Hardt 2014.

[263] Voir Royal Society 2017, p. 99.

Même parmi les systèmes d'IA prenant des décisions qui affectent des êtres humains, les risques varient en fonction du secteur, et les règles devraient varier aussi. Il est impossible de juger dans l'abstrait de l'équité des décisions d'IA. Dans chaque secteur, dans chaque domaine d'application, un même argument prend un poids différent[264]. Et dans chaque secteur s'appliquent des principes normatifs et juridiques différents. Par exemple, le droit à un procès équitable et la présomption d'innocence sont importants en droit pénal. En revanche, la liberté contractuelle est un principe important dans les transactions commerciales avec les consommateurs. Les nouvelles règles envisagées doivent donc couvrir des secteurs spécifiques.

Le besoin de nouvelles normes pourrait être évalué comme suit. Dans chaque secteur, il conviendrait de répondre à plusieurs questions :

(i) quelles règles s'appliquent dans ce secteur, avec quelle logique sous-jacente ? Une règle peut par exemple viser à protéger un droit de l'homme, exprimer un principe de droit (comme l'égalité, la liberté contractuelle, ou le droit à un procès équitable). Les impératifs économiques diffèrent aussi d'un secteur à l'autre : la mutualisation des risques est importante dans l'assurance, mais pas dans la plupart des autres secteurs. La justification sous-jacente des règles varie donc selon le secteur.

(ii) Comment est ou pourrait être utilisée la prise de décision automatisée dans le secteur concerné, et moyennant quels risques ? Un faux positif constitue un grave problème dans le système pénal : il peut conduire à l'interrogation, à l'arrestation, voire à la punition d'une personne. Nous ne devrions pas accepter un mode de décision d'IA qui porte atteinte aux valeurs fondamentales du droit pénal. En revanche, si une erreur d'IA fait payer davantage un consommateur par différenciation des prix, les conséquences sont souvent moins graves que celles de l'erreur de décision d'IA entraînant une arrestation.

(iii) Au vu de la logique sous-jacente des règles applicables au secteur concerné, faudrait-il revoir la réglementation pour tenir compte des décisions d'IA ? L'IA menace-t-elle les principes fondamentaux et les buts de la loi ? Si les textes en vigueur ne couvrent pas des risques importants, des modifications seraient à envisager.

En conclusion, les décisions d'IA pourraient appeler de nouvelles règles de protection de la justice et des droits de l'homme, comme le droit à la non-discrimination. Mais des recherches et des débats doivent encore être consacrés à la question de savoir si des règles sont nécessaires, et lesquelles.

### *Recherche empirique et technique*

Une bonne politique se fonde sur une bonne information. Il est clair qu'il faudrait en savoir davantage sur la discrimination provoquée par l'IA, et donc intensifier la recherche dans ce domaine[265]. Les États membres du Conseil de l'Europe devraient encourager la recherche des organes de surveillance du respect des droits de l'homme, des organismes de promotion de l'égalité et des universités. Il faudrait par exemple procéder à de nouvelles recherches empiriques. On ne distingue pas clairement à quelle échelle il est recouru aux décisions automatisées. Les décisions algorithmiques conduisent-elles fréquemment à des discriminations (sur la base de l'origine raciale, par exemple) et à d'autres formes de discrimination injustifiée ?

Il faudrait aussi consacrer davantage de recherche en informatique aux solutions. Comment pourrait, par exemple, être conçu un système d'IA qui respecte et promeuve les droits de l'homme, la justice et la transparence ? Est-il possible de vérifier que les données d'apprentissage ne vont pas produire de biais discriminatoires[266] ? Nous l'avons vu, ces questions constituent un domaine en plein essor dans les sciences informatiques[267]. D'une manière plus générale, dans les pays qui financent la recherche en IA, une partie des crédits devrait être affectée à la recherche sur les risques d'atteinte à la justice et aux droits de l'homme et sur leur réduction.

---

[264] Schauer 2003. Voir également Wachter et Mittelstadt 2018, p. 83.

[265] Voir Wagner *et al.* 2018, p. 43.

[266] Voir Campolo *et al.* 2018, p. 1

***Recherche normative et juridique***

Un débat public est également nécessaire, de même que des recherches normatives et juridiques. Comment serait-il possible d'imposer plus efficacement le respect de l'interdiction de la discrimination indirecte ? Comment la loi devrait-elle aborder la différenciation injustifiée qui échappe encore aux textes d'interdiction de la discrimination ? Comment définir la notion de justice selon le secteur ? Comment la loi — et la technologie — devraient-elles assurer la protection contre les aspects intersectionnels[268] et structurels de la discrimination[269] ? La loi devrait-elle protéger la confidentialité de certaines formes de « renseignements collectifs », et comment[270] ? Comment sauvegarder l'État de droit dans les décisions d'IA qui affectent des personnes[271] ? Certaines décisions ne devraient-elles jamais être laissées à un ordinateur, et lesquelles ? Comment le droit relatif à la protection des données pourrait-il être utilisé dans la pratique pour lutter contre la discrimination ? Faut-il fixer de nouvelles règles, ou suffirait-il d'ajuster les textes existants sur la discrimination et la protection des données ? Quels seraient les ajustements nécessaires ? Quelles seraient les nouvelles règles nécessaires ?

## VII. CONCLUSION

En conclusion, l'IA offre des possibilités nombreuses et précieuses pour améliorer nos sociétés. Mais ses décisions présentent aussi des risques : souvent opaques, elles peuvent avoir des effets discriminatoires — par exemple lorsque l'apprentissage du système se fonde sur des données reflétant des décisions humaines biaisées.

Les organisations publiques et privées peuvent recourir à des décisions d'IA lourdes de conséquences pour des personnes. Les entités publiques peuvent le faire à des fins de « police prédictive » ou pour recommander une peine, ou encore pour accorder ou refuser le versement de pensions de retraite, d'aides au logement ou d'allocations de chômage. Le privé peut aussi le faire d'une façon qui affecte gravement des personnes, comme en matière d'emploi, de logement ou de crédit. De plus, l'accumulation d'un grand nombre de décisions de moindre importance peut avoir des effets considérables. Une annonce publicitaire ciblée est rarement un problème majeur en soi, mais la publicité ciblée prise dans son ensemble peut exclure certains groupes. Et la différenciation des prix par l'IA pourrait faire que certains groupes paient systématiquement des prix supérieurs.

Les principaux instruments juridiques permettant de réduire les risques de discrimination par l'IA sont la réglementation anti-discrimination et la réglementation relative à la protection des données. Convenablement appliquées, toutes deux pourraient contribuer à la lutte contre la discrimination illicite. Les États membres du Conseil de l'Europe, les organes de contrôle du respect des droits de l'homme (comme la Commission européenne contre le racisme et l'intolérance) et les organismes de promotion de l'égalité devraient travailler à ce que le respect des normes actuelles de lutte contre la discrimination soit mieux garanti.

---

[267] https://www.fatml.org, consulté le 2 octobre 2018.

[268] Sur la discrimination intersectionnelle, voir Crenshaw 1989 ; Fredman 2016. Voir également ECRI Recommandation de politique générale nº 14 sur la lutte contre le racisme et la discrimination raciale dans le monde du travail, adoptée le 22 juin 2012, CRI(2012)48, https://rm.coe.int/ecri-general-policy-recommendation-no-14-on-combating-racism-and-racia/16808b5afc (consulté le 14 octobre 2018).

[269] Sur la discrimination structurelle, voir le paragraphe 20 de l'exposé des motifs de la Recommandation de politique générale nº 2 (2018) de l'ECRI.

[270] Voir Bygrave 2002, chapitre 9-16 ; Taylor, Van der Sloot, et Floridi 2017 ; Vedder 1997.

[271] Hildebrandt 2015 ; Bayamlıoğlu et Leenes 2018.lo

Mais l’IA permet aussi d’opérer de nouvelles formes de différenciations injustifiées (ou de discrimination) que n’envisagent pas les textes en vigueur. La plupart des lois anti-discrimination ne s’appliquent qu’à la discrimination fondée sur des caractéristiques protégées, comme l’origine raciale. Elles ne couvrent pas les organisations qui pratiquent la différenciation sur des catégories nouvelles, non corrélées avec des caractéristiques protégées. Cette différenciation n’en pourrait pas moins être injuste, par exemple lorsqu’elle renforce les inégalités sociales. De nouvelles normes seront très probablement nécessaires pour protéger la justice et les droits de l’homme dans le domaine de l’IA. Mais réglementer l’ensemble de cette dernière ne serait pas une bonne façon de faire, car les utilisations des systèmes d’IA sont trop diverses pour un seul dispositif réglementaire. Il faudra concevoir des règles sectorielles, du fait que les valeurs sous-jacentes et les problèmes varient d’un secteur à l’autre. Des débats et des recherches interdisciplinaires sont encore nécessaires. Si nous faisons aujourd’hui les bons choix, nous pourrons bénéficier de tous les avantages de l’IA, tout en minimisant les risques de discrimination injustifiée.

## BIBLIOGRAPHIE